\documentclass[11pt, a4paper]{article}

\usepackage{amsmath,amssymb,amsthm}
\usepackage{mathtools}
\usepackage{graphicx}
\usepackage[font=small,labelfont=bf]{caption, subcaption}
\usepackage{fullpage}
\usepackage{color}
\usepackage[colorlinks=true, citecolor=blue]{hyperref}
\usepackage[usenames,dvipsnames]{xcolor}
\usepackage{tikz}
\usepackage{authblk}
\usepackage{float}

\usepackage{algorithm}
\usepackage{algpseudocode}

\RequirePackage[capitalize,nameinlink,noabbrev]{cleveref}

\usepackage[T1]{fontenc}
\usepackage{libertinus}

\usepackage{authblk}

\newcommand{\nrb}[1]{{\color{red}\textbf{#1}}}

\usepackage{lipsum}

\usepackage{enumitem}

\newtheorem{conj}{Conjecture}

\theoremstyle{remark}
\newtheorem*{rem}{Remark}

\newcommand{\Nmin}{N_\text{min}}
\newcommand{\Nmax}{N_\text{max}}

\newcommand{\nmin}{n_\text{min}}

\newcommand{\ult}{\underline{\theta}}
\newcommand{\olt}{\overline{\theta}}

\numberwithin{equation}{section}

\title{Random sampling of self-avoiding theta-graphs}

\author[a,*]{Nicholas R. Beaton}
\author[ ]{Aleksander L. Owczarek}

\affil[a]{School of Mathematics and Statistics, University of Melbourne, Melbourne 3010, Australia}

\affil[*]{Corresponding Author:
\href{mailto:nrbeaton@unimelb.edu.au}{nrbeaton@unimelb.edu.au}}

\usepackage[giveninits, maxbibnames=3, sorting=nyt, sortcites=true, style=numeric-comp, abbreviate=false]{biblatex}
\renewbibmacro{in:}{}
\DeclareFieldFormat[misc]{title}{\mkbibquote{#1}}

\DeclareFieldFormat[article]{volume}{\mkbibbold{#1}}
\DeclareFieldFormat{url}{\textsc{url}: \href{#1}{#1}}
\DeclareFieldFormat{doi}{\textsc{doi}: \href{http://dx.doi.org/#1}{#1}}
\DeclareFieldFormat{eprint:arxiv}{arXiv:\href{https://arxiv.org/abs/#1}{#1}}
\begin{document}

\maketitle

\begin{abstract}
    Theta-graphs are a type of spatial graph with two vertices connected by three edges. We investigate embeddings of theta-graphs in the square and simple cubic lattices, using a combination of the Wang-Landau Monte Carlo method with a variant of the BFACF algorithm which accommodates vertices of degree 3. This allows us to estimate the critical exponents governing the number of theta-graphs and the distributions of the different arm-lengths. For the cubic lattice these values can be compared to the corresponding exponents for prime knots. We also study the number of `monodisperse' theta-graphs where the three arms have the same lengths, and find evidence supporting a conjecture for the critical exponent in two dimensions.
\end{abstract}

\section{Introduction}\label{sec:introduction}

Self-avoiding walks (SAWs) and polygons (SAPs) on regular lattices are well known models of linear and ring polymers \cite{madras_self-avoiding_1996,DeGennes1979,janse_van_rensburg_statistical_2015}. In three dimensions, in particular, SAWs and SAPs display properties similar to those of real world polymers in a good solvent, such as the critical exponents which govern geometric quantities \cite{madras_self-avoiding_1996,DeGennes1979,janse_van_rensburg_statistical_2015}.

\begin{figure}
    \centering
    \begin{subfigure}{0.19\textwidth}
    \centering
    \includegraphics[height=3cm]{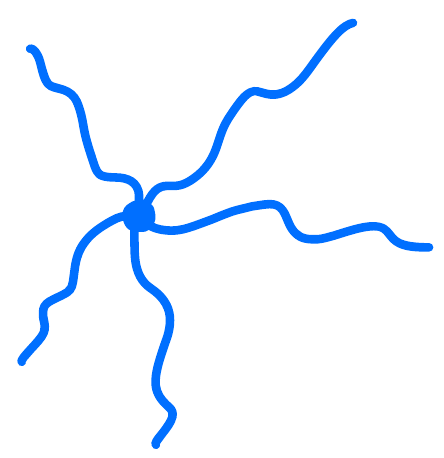}
    \caption{}
    \end{subfigure}
    \hfill
    \begin{subfigure}{0.19\textwidth}
    \centering
    \includegraphics[height=3cm]{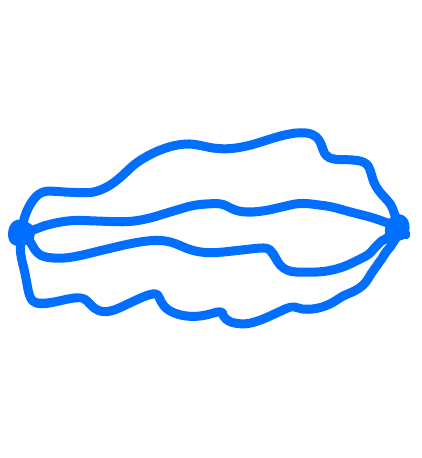}
    \caption{}
    \end{subfigure}
    \hfill
    \begin{subfigure}{0.2\textwidth}
    \centering
    \includegraphics[height=3cm]{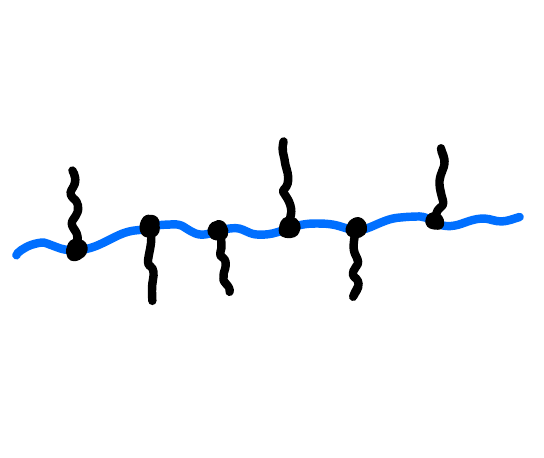}
    \caption{}
    \end{subfigure}
    \hspace{1em}
    \hfill
    \begin{subfigure}{0.2\textwidth}
    \centering
    \includegraphics[height=3cm]{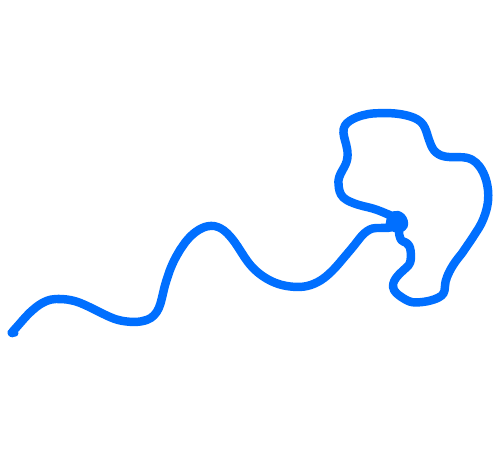}
    \caption{}
    \end{subfigure}
    \hfill
    \begin{subfigure}{0.15\textwidth}
    \centering
    \includegraphics[height=3cm]{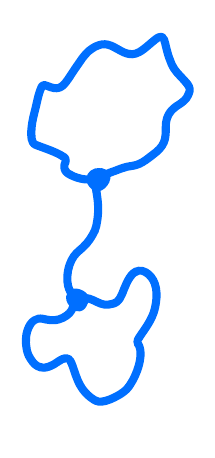}
    \caption{}
    \end{subfigure}
    \caption{\textbf{(a)} A star, \textbf{(b)} a watermelon, \textbf{(c)} a comb, \textbf{(d)} a tadpole, and \textbf{(e)} a dumbbell.}
    \label{fig:various_objects}
\end{figure}

In recent years, lattice models of more complicated topologies than linear and ring polymers have received attention. Ring polymers themselves in three dimensions can have non trivial knot type and so the study of the behaviour lattice polygons regarding knot type have been a focus. 
Another way to generalise the topology of lattice objects (and so the consideration of more complicated polymers) is to allow vertices of degree 3 or more. This can be done in a variety of ways, giving structures like stars \cite{bradly_force-induced_2019}, watermelons \cite{duplantier_polymer_1986,duplantier_statistical_1989}, combs \cite{janse_van_rensburg_exponential_2024}, tadpoles \cite{guttmann_limiting_1973}, theta-graphs \cite{no_topological_2021,kim_lattice_2022} and dumbbells \cite{guttmann_two-dimensional_1978}. See \cref{fig:various_objects} for some illustrations.

\begin{figure}
    \centering
    \includegraphics[width=0.45\textwidth]{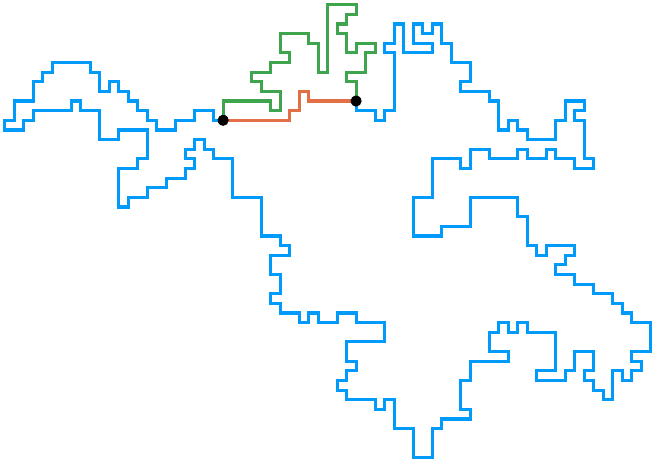}
    \caption{A size 500 theta on the square lattice, with arm lengths 18, 66 and 416.}
    \label{fig:size_500_theta}
\end{figure}

\begin{figure}
    \centering
    \begin{subfigure}{0.2\textwidth}
    \centering
        \includegraphics[height=3cm]{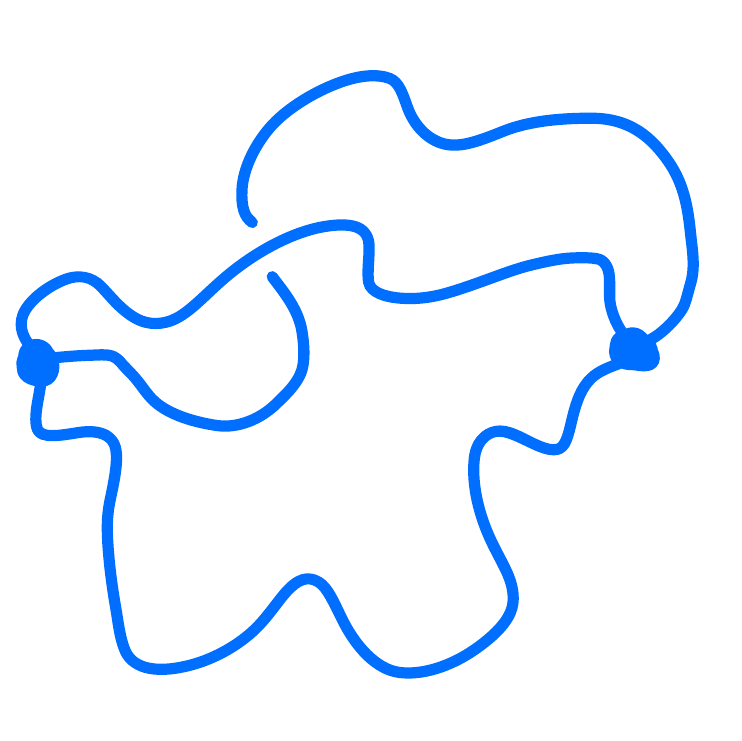}
        \caption{}
    \end{subfigure}
    \begin{subfigure}{0.2\textwidth}
    \centering
        \includegraphics[height=3cm]{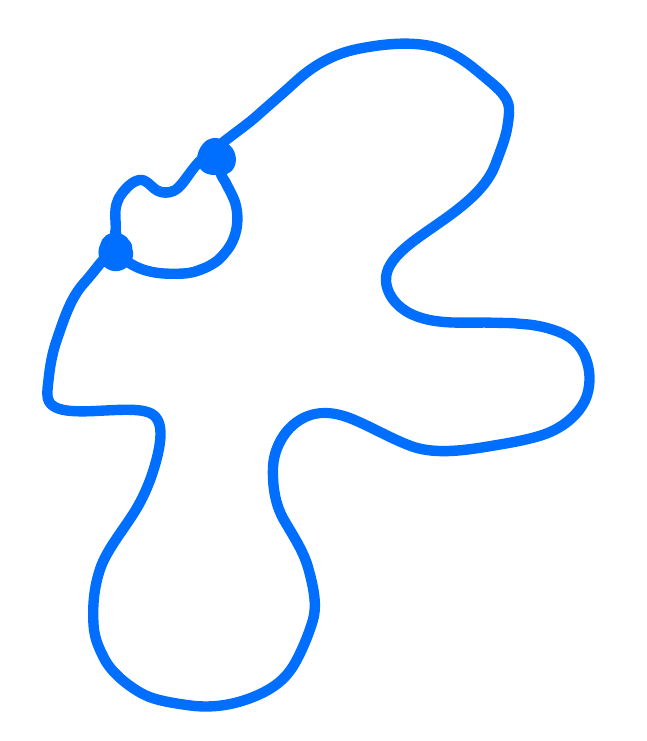}
        \caption{}
    \end{subfigure}
    \begin{subfigure}{0.2\textwidth}
        \centering
        \includegraphics[height=3cm]{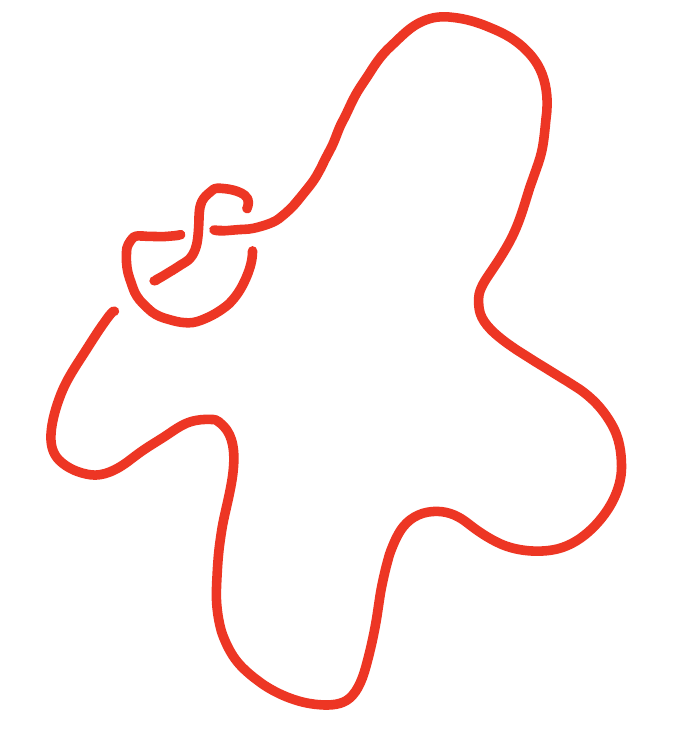}
        \caption{}
    \end{subfigure}
    \begin{subfigure}{0.2\textwidth}
    \centering
        \includegraphics[height=3cm]{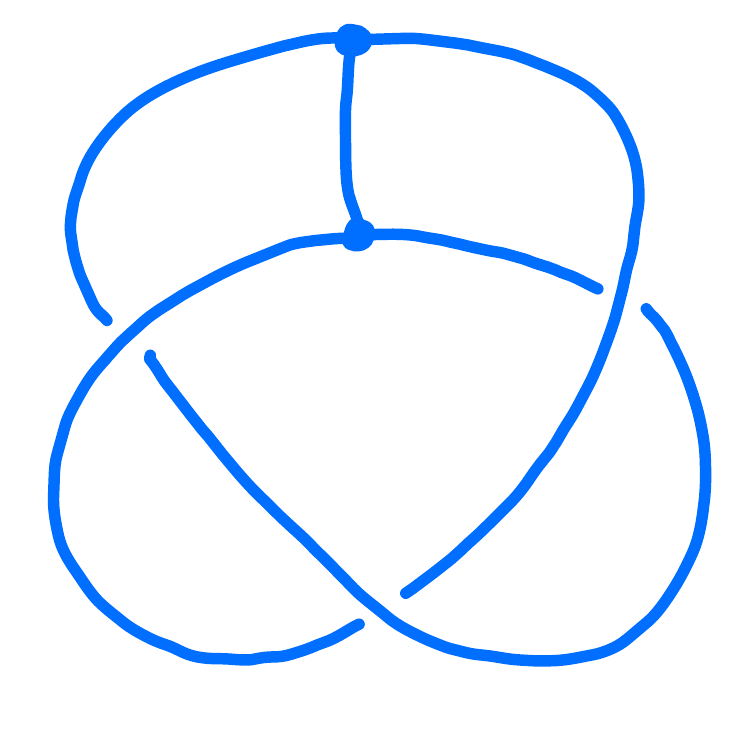}
        \caption{}
    \end{subfigure}
    \caption{\textbf{(a)} and \textbf{(b)} Two thetas with different distributions of arm lengths. \textbf{(c)} A trefoil knot with a localised knot component. \textbf{(d)} A theta with non-trivial `knot' type.}
    \label{fig:thetas_knot_illustration}
\end{figure}

In this paper, we focus on theta-graphs (\emph{thetas} for short). As graphs, these comprise two vertices and three edges, with each edge connecting one vertex to the other. In three dimensions thetas can have different `knot' types, which can be grouped according to their crossing number in a similar manner to regular knots, and for which there exist invariants for distinguishing them \cite{moriuchi_enumeration_2009}. While the methods we use here can be applied to thetas of any specified knot type, we focus only on `unknotted' thetas with crossing number 0 (equivalently, those which can be embedded in a 2-sphere in $S^3$). This restriction does have some numerical consequences, which we discuss further in \cref{ssec:enumeration}. See \cref{fig:thetas_knot_illustration} for some schematics, and \cref{fig:size_500_theta} for a example of a square lattice theta.

Theta-shaped polymers have been synthesised in laboratories \cite{tezuka_synthesis_2003}, and proteins with a theta topology have been observed \cite{dabrowski-tumanski_theta-curves_2024}. Thetas are one type of a more general class called polymer networks, which characterise important materials including gels and rubbers \cite{gu_polymer_networks_2020}. Thetas formed by Gaussian chains (rather than lattice self-avoiding walks) were studied in \cite{uehara_statistical_2018}.

Our initial motivation for this particular study was twofold. Firstly, two-dimensional lattice thetas are one type of object considered by Duplantier \cite{duplantier_polymer_1986,duplantier_statistical_1989}, who conjectured (among other things) the (entropic) critical exponent $\gamma$ for polymer networks of fixed topology. In particular, these networks are `monodisperse', i.e.\ each `arm' connecting a pair of vertices has the same length. In this paper we study thetas where the three arms can have varying length (`polydisperse'), but by also considering the monodisperse subset we have been able to check the validity of Duplantier's conjecture (see \cref{sec:equilateral}).

Our second motivation is to compare the typical `shape' of thetas with that of typical knots. Long ring polymers are knotted with high probability \cite{frisch_chemical_1961,delbruck_mathematical_1962} and there has been considerable interest in how the topology of polymers like DNA affect biological function (see e.g.\ the review \cite{tubiana_topology_2024}). One question of interest is whether the `knotted part' of a typical large prime knot is localised in a small region of the ring, or if it is distributed around the whole structure. Various numerical experiments  have shown that knots tend to be `weakly localised', with the average size of the knotted part of a ring polymer of size $n$ scaling like $n^t$. Estimates for $t$ have ranged from $0.4$ \cite{farago_pulling_2002}, to $0.65$ \cite{virnau_knots_2005,mansfield_properties_2010} and $0.75$ \cite{marcone_size_2007,orlandini_size_2009}. Numerical experiments with linear chains \cite{tubiana_spontaneous_2013} have also found power-law behaviour, with an exponent around $0.44$.  We are interested in whether a typical theta looks, in some sense, like a typical knot, with the two shorter arms of a theta comprising a small `bubble' within a large polygon. See \cref{fig:thetas_knot_illustration} for an illustration.

While exact enumeration and series analysis methods for SAWs, SAPs and related objects have been very productive in two dimensions \cite{guttmann_polygons_2009}, they have generally had limited success in three (or more) dimensions. In these cases Monte Carlo methods have been more effective, and many different algorithms have been developed over the past few decades. These include Beretti-Sokal, the pivot algorithm, PERM, GARM, and GAS. (See the review \cite{janse_van_rensburg_monte_2009} for descriptions and references for all of these.)

In this paper we combine local BFACF-type moves \cite{Arag_o_de_Carvalho_1983,de_Carvalho_1983,Berg_1981} with the Wang-Landau algorithm \cite{wang_landau_2001}. This enables us to sample thetas of fixed knot type (we have focussed on unknots) across a range of shapes and sizes, in order to compute estimates of the number of thetas of a given size, as well as the distribution of the length of shortest or second-shortest arm, or the sum of both. From these data, we calculate estimates for several critical exponents, namely those which govern the number of thetas, the number of monodisperse thetas, the average size of the shortest arm (or second-shortest, or the sum of both), and the mean squared distance between the two branch points.

The structure of the paper is as follows. In \cref{sec:bfacf} we describe the BFACF algorithm as well as its generalisation to thetas and other branching structures. In \cref{sec:wanglandau} we outline the Wang-Landau algorithm and its use for approximate enumeration. In \cref{sec:polydisperse} we present results for polydisperse theta graphs, including enumerative results and estimates for the distribution of arm-lengths and the separation of the branch points. In \cref{sec:equilateral} we present results for monodisperse thetas. Finally in \cref{sec:conclusion} we discuss some future avenues for research.

\section{The BFACF algorithm for theta-graphs}\label{sec:bfacf}

The BFACF algorithm \cite{Arag_o_de_Carvalho_1983,de_Carvalho_1983,Berg_1981} is an algorithm for sampling self-avoiding polygons or self-avoiding walks with fixed end-points. It is known to be ergodic for SAPs in the two-dimensional square or triangular lattices; that is, any polygon can be obtained from any other polygon by a sequence of BFACF moves. In the three-dimensional simple cubic, face-centred cubic (FCC) or body-centred cubic (BCC) lattices, the ergodicity classes for polygons are the knot types \cite{janse_van_rensburg_bfacf_1991,janse_van_rensburg_bfacf-style_2011}. 
In this work we will focus on the square and cubic lattices; see \cref{sec:conclusion} for some discussion of other lattices.

\begin{figure}
    \centering
        \includegraphics[width=0.4\textwidth]{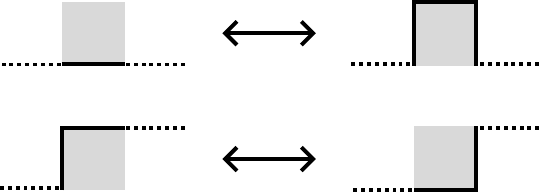}
        \caption{The basic BFACF moves for the square and cubic lattices.}
    
    \label{fig:basic_BFACF_moves}
\end{figure}

The elementary BFACF moves for polygons on the square and cubic are illustrated in \cref{fig:basic_BFACF_moves}. Note that on the square and cubic lattices, the BFACF moves either leave the total length unchanged, or modify it by $\pm2$.
In \cref{alg:bfacf_polygons} we outline the basic procedure for BFACF moves, as applied to SAPs.

\begin{algorithm}[ht]
    \caption{(BFACF move for SAPs)}
    \label{alg:bfacf_polygons}
    Let $P$ be a SAP defined by the set of edges $E=\{e_1,\dots,e_n\}$.
    \begin{algorithmic}[0] 
        \Procedure{BFACF}{$P$}
            \State $e \gets$ uniform sample from $E$
            \State $q \gets$ uniformly chosen plaquette adjacent to $e$
            \State $Q \gets$ set of lattice edges adjacent to $q$
            \State $E' \gets E \sqcup Q$ (i.e.\ $(E \cup Q) \setminus (E \cap Q)$)
            \If{$E'$ forms the edges of a SAP}
                \State $P' \gets E'$
            \ElsIf{$E'$ does not form a SAP}
                \State $P' \gets P$
            \EndIf
            \State \Return{$P'$}
        \EndProcedure
    \end{algorithmic}
\end{algorithm}

BFACF moves typically form part of the implementation of a Markov chain Monte Carlo (MCMC) study, where the probability of accepting or rejecting a given move is determined by another algorithm. Examples include use with the Metropolis algorithm \cite{rensburg_dimensions_1991} and with GAS (generalised atmospheric sampling) \cite{janse_van_rensburg_generalized_2012}.

\begin{figure}
    \centering
    \begin{subfigure}{0.7\textwidth}
        \begin{tikzpicture}
            \node at (0,0) [anchor=south west] {\includegraphics[width=\textwidth]{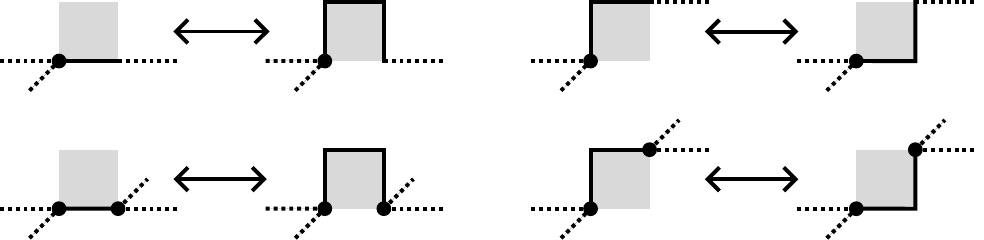}};
            \node at (2.7,2) {$2_1$};
            \node at (8.7,2) {$0_1$};
            \node at (2.7,0.3) {$2_2$};
            \node at (8.7,0.3) {$0_2$};
        \end{tikzpicture}
        \caption{}
        \label{fig:sq_BFACF_theta_fixtri}
    \end{subfigure}

    \vspace{2ex}
    \begin{subfigure}{0.7\textwidth}
        \begin{tikzpicture}
            \node at (0,0) [anchor=south west] {\includegraphics[width=\textwidth]{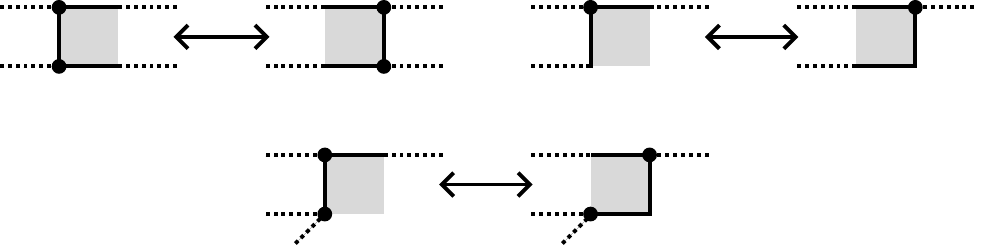}};
            \node at (2.7,2) {$0_3$};
            \node at (8.7,2) {$1_1$};
            \node at (5.7,0.3) {$1_2$};
        \end{tikzpicture}
        \caption{}
        \label{fig:sq_BFACF_theta_movetri}
    \end{subfigure}
    \caption{The set of BFACF moves for spatial graphs on the cubic lattice. The moves in \textbf{(a)} do not change the location of the vertices of degree 3, while the moves in \textbf{(b)} do move at least one of the degree 3 vertices. These are in addition to the regular BFACF moves in \cref{fig:basic_BFACF_moves} (a). This figure is adapted from \cite[Fig.~4.1]{Tamaki2018}.}
    \label{fig:sq_BFACF_theta}
\end{figure}
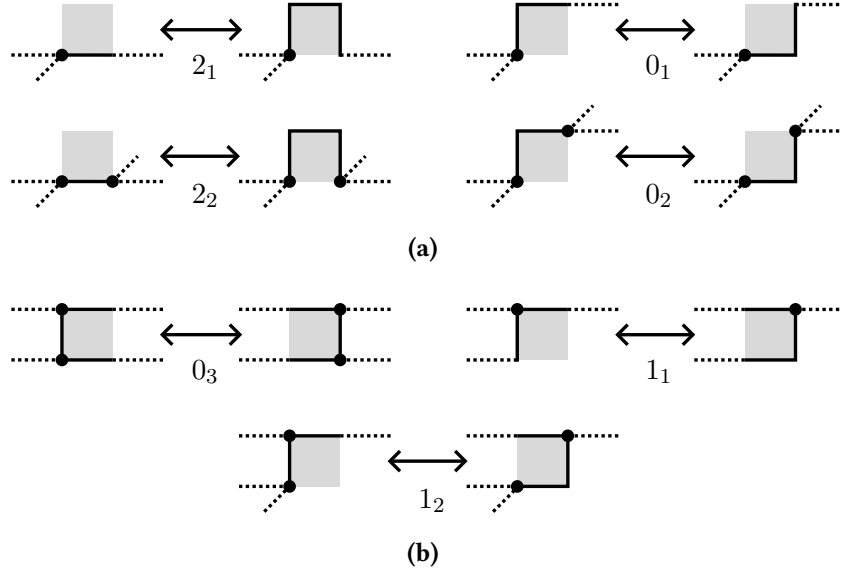

Tamaki~\cite{Tamaki2018} found a set of BFACF-type moves for spatial graphs (i.e.\ graphs embedded in space) in the cubic lattice with vertices of degree 2 and 3. In particular, they found that the ergodicity classes are the graph types (generalising the idea of knot types, i.e.\ two graphs have the same type if they are equivalent under ambient isotopy). The set of required moves is illustrated in \cref{fig:sq_BFACF_theta}. Note that some of these moves change the overall length by $\pm1$. It is this expanded set of moves which we implement.

While Tamaki did not specifically address the two-dimensional square lattice, we expect the ergodicity classes in the square lattice to also be the graph types.




\section{The Wang-Landau method for approximate enumeration}\label{sec:wanglandau}

The Wang-Landau method~\cite{wang_landau_2001,landau_new_2004} is a Monte Carlo method for estimating the density of states $g$ of a system, defined as
\begin{equation}
    g(E) := \frac{\mathrm{d}\Omega(E)}{\mathrm{d}E},
\end{equation}
where $\Omega(E)$ is the number of states with energy less than or equal to $E$. Equivalently, the number of states with energy in the small interval $[E,E+\delta E]$ is $N(E) = g(E)\delta E$. 

The idea is to perform a random walk in energy space. If $s_1$ and $s_2$ are two states with energy levels $E_1$ and $E_2$, let $p(s_2 |s_1)$ be the probability of proposing a move to $s_2$ given the current state $s_1$. Then the probability of accepting such a move has a Metropolis-Hastings form:
\begin{equation}
    P(E_1 \to E_2) = \min\left\{ \frac{g(E_1)p(s_1  | s_2)}{g(E_2)p(s_2 | s_1)}, 1\right\}.
\end{equation}
If the move is accepted then we update $g(E_2) \mapsto f\times g(E_2)$ where $f>1$ is a modification factor; otherwise we update $g(E_1)$ by the same factor. We also maintain a histogram $H$ which tracks the number of visits to each energy level: after the aforementioned accepted/rejected move, we increment either $H(E_1)$ or $H(E_2)$ by 1 as appropriate. In practice the values $g(E)$ become large very quickly, so instead we record $\log g(E)$ and update via $\log g(E) \mapsto \log g(E) + \log f$. A typical initial choice for $f$ is $f_0 = e^1$. We will follow \cite{gleeson_thesis_2024} and sometimes refer to the factor $\frac{p(s_1  | s_2)}{p(s_2 | s_1)}$ as the \emph{Hastings factor} (for many systems, this factor is just 1).

After this algorithm has run sufficiently long, the histogram $H$ should be approximately flat, i.e.\ all energy levels have been visited approximately equally often. A simple threshold is $\frac{\min(H)}{\max(H)} \geq X$ for some fraction $X$; in this paper we use $X = 90\%$. (One must choose some frequency with which to check for the flatness of $H$; typically every $M$ moves for some $M$ which depends on the system size.) At this point $g$ will have converged to its true value, within an accuracy proportional to $\log f$. We then decrease $f$ via some function (often $f \mapsto \sqrt{f}$), reset $H=0$, and then start again.

This process repeats until some criteria has been met: either $f < 1+\epsilon$ for a small $\epsilon$, or a sufficient (large) number of moves have been made. At this point, $g(E)$ provides an estimate for the \emph{relative} density of states. It must be normalised, typically by knowledge of the actual number $g_\text{ex}(E_0)$ of ground states with energy $E_0$. Then one scales all values by
\begin{equation}\label{eqn:wl_final_rescaling}
\log g(E) \mapsto \log g(E) - \log g(E_0) + \log g_\text{ex}(E_0)
\end{equation}
to obtain the final estimate.

This method works well up to a certain accuracy, but as $f$ continues to decrease a problem can appear. Because we decrease $f$ by $f \mapsto \sqrt{f}$, the value $\log f$ with which we are updating $g(E)$ decreases exponentially. This turns out to be too fast -- with $\log f$ decreasing exponentially, the estimates $g(E)$ essentially converge and get `frozen', with no further improvements to accuracy. See \cref{fig:saps_errors} for a plot which demonstrates this in the case of SAPs.

A solution to this problem was proposed in \cite{belardinelli_fast_2007}. The idea is to decrease $F = \log f$ more slowly: instead of updating $F \mapsto F/2$ when the histogram $H$ is sufficiently flat, we use $F = 1/t$, where $t$ is the Monte Carlo time (proportional to the number of proposed moves). The method given in \cite{belardinelli_fast_2007} (and which we use here) is actually two-stage: the exponential updating $F \mapsto F/2$ is used at first, until $F < 1/t$. At that point, we switch to $F=1/t$, and the histogram $H$ is no longer used. In practice $F$ is still only updated periodically, every $M$ moves for some fixed $M$. With the error at energy level $E$ defined as \cite{belardinelli_fast_2007}
\begin{equation}\label{eqn:wl_error}
    \epsilon(E) = \left|1-\frac{\log g(E)}{\log g_\text{ex}(E)}\right|,
\end{equation}
it is then expected \cite{belardinelli_fast_2007} that the average error $\langle \epsilon \rangle$ (across all energy levels) should scale as $\sqrt{F} = t^{-1/2}$.

The Wang-Landau method has seen a wide variety of applications, including analysis of computer networks \cite{Atisattapong_2021}, numerical integration \cite{LI2007524}, interacting polymers \cite{taylor_phase_2009,wust_versatile_2009} and the Ising model \cite{ZHAN2008339}. The method has been adapted in a variety of ways, including parallel implementations \cite{ZHAN2008339,YIN20121568} and in combination with the $N$-fold way \cite{malakis_wanglandau_2004}.

\subsection{General framework for enumeration}

To apply the Wang-Landau method to the approximate enumeration of combinatorial objects like self-avoiding polygons, we need only make a few small changes. Some of the ideas in this and the following subsections were presented in \cite{gleeson_thesis_2024}.

First, let $\mathcal{C}$ be a combinatorial class with size function $|\cdot| : \mathcal{C} \to \mathbb{N}_0$. Let $\mathcal{C}_n$ be the set of objects of size $n$, and define $C_n = |\mathcal{C}_n| < \infty$. (Note that here we use $|\cdot|$ to denote both the size function on $\mathcal{C}$ as well as the usual cardinality of a set.)

Let $\Nmin$ and $\Nmax$ be respectively the minimum and maximum sizes of the objects we wish to count. Then the set $\mathcal{S}$ of states is the set of objects with size in the interval $I = [\Nmin, \Nmax]$. The `energy' of an object is its size. Let $B = |\{n \in I : C_n > 0\}|$.

Define a probability distribution $\rho$ on $\mathcal{S}$ where $\rho(\gamma)$ depends only on $|\gamma|$. That is, $\rho(\gamma) = R(|\gamma|)$ where $R : I \to [0,1]$. Then writing $G(n) = \log g(n)$,
\begin{equation}
    \frac{1}{R(n)} = \exp(G(n))
\end{equation}
now becomes an estimate of the relative multiplicity of $\mathcal{C}_n$. If we know $C_m$ for some $m \in I$, then the multiplicities of the other $\mathcal{C}_n$ can be estimated by
\begin{equation}
    \tilde{C}_n = C_m \exp(G(n) - G(m)),
\end{equation}
which is equivalent to \eqref{eqn:wl_final_rescaling}.
As with the original Wang-Landau method, we require an irreducible random walk over the objects in $\mathcal{S}$. Then the probability distribution $\rho$ is regularly updated (via the updating of $G$) according to which states are sampled. The general framework is given in \cref{alg:basic_WL_enum}.

\begin{algorithm}[ht]
    \caption{(Basic Wang-Landau for enumeration)}
    \label{alg:basic_WL_enum}
    Let $\mathcal{C}$ and $I$ be as above. Suppose $m\in I$ is such that $C_m$ is known. Let $\textsc{MetropolisHastings}(s,\rho)$ denote a sample from the distribution $\rho$ given the previous state $s$. The Wang-Landau algorithm for estimating the quantities $C_n$ is:
    \begin{algorithmic}[0] 
        \Procedure{WangLandau}{}
            \State $s \gets$ initial state
            \State $G \gets [0,\dots,0]$
            \State $H \gets [0,\dots,0]$
            \State $F \gets 1$
            \State $i = 0$
            \While{$i < M$} \Comment{$M$ is the total number of samples to take}
                \State $s \gets \textsc{MetropolisHastings}(s,\exp(-G))$
                \State $G[s] \gets G[s] + F$
                \State $H[s] \gets H[s]+1$
                \State $i \gets i+1$
                \If{$H$ sufficiently flat}
                    \State $H \gets [0,\dots,0]$
                    \State $F \gets F/2$
                \EndIf
            \EndWhile
            \State $\tilde C \gets C_m \exp(G - G(m))$
        \EndProcedure
    \end{algorithmic}
\end{algorithm}
Note that instead of setting $M$ to be the total number of samples in \cref{alg:basic_WL_enum}, we can instead set a minimum value $\epsilon$ for $F$, replacing the \textbf{while} $i < M$ \textbf{do} loop with \textbf{while} $F > \epsilon$ \textbf{do}.

To avoid saturation of errors we can improve this using the $1/t$ algorithm, given in \cref{alg:WL_enum_1t}. Again the condition $i<M$ can be replaced by $F > \epsilon$ for some small $\epsilon$.

\begin{algorithm}
    \caption{(Improved Wang-Landau for enumeration)}
    \label{alg:WL_enum_1t}
    \begin{algorithmic}[0] 
        \Procedure{ImprovedWangLandau}{}
            \State $s \gets$ initial state
            \State $G \gets [0,\dots,0]$
            \State $H \gets [0,\dots,0]$
            \State $F \gets 1$
            \State $i \gets 0$
            \State \textit{stage} $\gets 1$
            \While{$i < M$}
                \State $s \gets \textsc{MetropolisHastings}(s,\exp(-G))$
                \State $G[s] \gets G[s] + F$
                \State $H[s] \gets H[s]+1$
                \State $i \gets i+1$
                \If{\textit{stage} $=1$ and $H$ sufficiently flat}
                    \State $H \gets [0,\dots,0]$
                    \State $F \gets F/2$
                    \If{$f < B/i$}
                        \State $F \gets B/i$
                        \State \textit{stage} $\gets 2$
                    \EndIf
                \ElsIf{\textit{stage} $=2$}
                    \State $F\gets B/i$
                \EndIf
            \EndWhile
            \State $\tilde C \gets C_m \exp(G - G(m))$
        \EndProcedure
    \end{algorithmic}
\end{algorithm}

\begin{figure}
    \centering
    \includegraphics[width=0.55\textwidth]{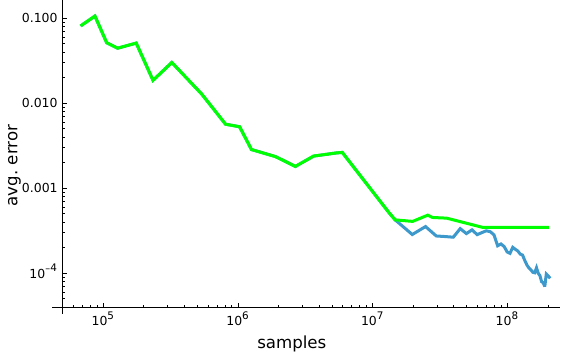}
    \caption{A plot of the average error \eqref{eqn:max_rel_error} for the Wang-Landau algorithm, sampling SAPs on the square lattice up to size $n_\text{max} = 130$, with a total of $2\times10^8$ samples. \cref{alg:basic_WL_enum} is in green, while \cref{alg:WL_enum_1t} coincides with \cref{alg:basic_WL_enum} up to $\approx 1.47\times10^7$ samples, after which it switches to the ``$1/t$ algorithm'' and is plotted in blue.}
    \label{fig:saps_errors}
\end{figure}

In \cref{fig:saps_errors} we compare \cref{alg:basic_WL_enum,alg:WL_enum_1t} for SAPs on the square lattice. The exact counts $p_n$ are known up to size $N_\text{max}=130$ \cite{clisby_new_2012}. We ran the Wang-Landau algorithm on SAPs (combined with BFACF moves, see \cref{ssec:SAPs_overview}) up to size $N_\text{max}$ until $2\times10^8$ samples have been taken, periodically computing estimates $\tilde{p}_n$ from $G$, and computing the average error $\langle \delta \rangle$, where
\begin{equation}\label{eqn:max_rel_error}
    \delta(n) = \left| 1-\frac{\log \tilde{p}_n}{\log p_n}\right|
\end{equation}
as per \eqref{eqn:wl_error}.
The saturation of errors in \cref{alg:basic_WL_enum} can clearly be seen in \cref{fig:saps_errors}, where after some time $f$ becomes so small that $G$ (and hence the estimates $\tilde{p}_n$) has essentially converged. After this point there is nothing to be achieved by running the algorithm any longer.

\subsection{Self-avoiding polygons}\label{ssec:SAPs_overview}

To approximately enumerate (unknotted) self-avoiding polygons of length $n$ we can use \cref{alg:basic_WL_enum} or \cref{alg:WL_enum_1t}. On the square and cubic lattices we have $\Nmin=4$.
Then
\begin{equation}
    C_4 = \begin{cases} 1, & \text{square} \\ 3, & \text{cubic.} \end{cases} 
\end{equation}

It remains to define the random walk over the set of SAPs of size $\leq \Nmax$. This is where the BFACF algorithm comes in (see \cref{alg:bfacf_polygons}).

Note that when implementing BFACF moves in Wang-Landau, care must be taken with the length-changing moves: there is only one way to choose an edge and adjacent plaquette when increasing the length, but there are multiple ways to choose an edge/plaquette when decreasing the length. When using Wang-Landau this difference can be accounted for in the Hastings factor. For the square and cubic lattices, we have (when $|S| = n$)
\begin{equation}\label{eqn:Hastings_factors}
    \frac{p(S | S')}{p(S'|S)} = \begin{cases} \frac{3n}{n+2} & \text{if increasing length by 2} \\
    \frac{n}{3(n-2)} & \text{if decreasing length by 2} \\
    1 & \text{if length is unchanged.}
    \end{cases}
\end{equation}
An alternative strategy is to restrict the types of length-decreasing moves, so that they are exactly in one-to-one correspondence with length-increasing moves. This simplifies the Hastings factors slightly (it eliminates the factors of 3 and 2 in the numerators and denominators) but does mean that certain choices of $e$ and $q$ in \cref{alg:bfacf_polygons} will be immediately rejected.

\subsection{Spatial graphs}

\cref{alg:bfacf_polygons} can be generalised to spatial graphs with vertices of degree 3 in a straightforward manner. 
One still chooses an edge $e$ and an adjacent plaquette $q$ uniformly at random.
\begin{itemize}
    \item If $q$ is not incident on any vertex of degree 3, then proceed as in \cref{alg:bfacf_polygons}. The Hastings factor is the same as for SAPs.
    \item If $q$ is incident on one or two vertices of degree 3, then one of the moves illustrated in \cref{fig:sq_BFACF_theta} will be proposed. The moves in \cref{fig:sq_BFACF_theta} (a) do not shift a vertex of degree 3 and thus the updating is the same as in \cref{alg:bfacf_polygons} (again the Hastings factor is the same as for SAPs). The moves in \cref{fig:sq_BFACF_theta} (b) do shift one or more vertex of degree 3, and hence the updating is not as simple as `inverting' the edges around the plaquette. The Hastings factor is still straightforward to compute.
    \item If $q$ is incident on three or more vertices of degree 3 (this is of course not possible with thetas), no move is proposed.
\end{itemize}



\section{Results: polydisperse theta-graphs}\label{sec:polydisperse}

For our purposes a \emph{theta-graph} (\emph{theta} for short) is a spatial graph with two vertices of degree 3, which we denote $v_1$ and $v_2$, arranged so that there are three ``arms'' connecting $v_1$ and $v_2$. In three dimensions there are actually infinitely many topologically distinct spatial graphs which satisfy this property, just as there are infinitely many knot types for a simple closed curve. See for example \cite{moriuchi_enumeration_2009}. Because the BFACF algorithm preserves spatial graph type, in this work we restrict to ``unknotted'' thetas, as per \cref{fig:thetas_knot_illustration}. However, we do note that different types of thetas can also be sampled using the BFACF algorithm. We expect that thetas with other ``knot'' types  will have different critical exponents, and we are not sure what other kind of quantitative differences one might expect to find. For the remainder of the paper we will just use ``theta'' to refer to unknotted theta-graphs. Let $\mathcal{T}_n$ be the set of theta-graphs with $n$ edges.

On a given lattice we say that a theta is \emph{polydisperse} if the three arms connecting $v_1$ and $v_2$ may have different lengths. If instead we restrict to thetas where all three arms have the same length, we say such objects are \emph{monodisperse} (see \cref{sec:equilateral}).

For each of the square and cubic lattices we ran the Wang-Landau algorithm in three different ways:
\begin{enumerate}[label=(\Roman*)]
    \item flattening across the size $n$ and the length $\ell_1$ of the shortest arm; \label{itemI}
    \item flattening across the size $n$ and the length $\ell_2$ of the second-shortest (equivalently, second-longest) arm; \label{itemII}
    \item flattening across the size $n$ and the sum $\ell_1+\ell_2$ of the shortest and second-shortest arms. \label{itemIII}
\end{enumerate}
For \ref{itemI}--\ref{itemIII}  we sampled thetas up to size 500. For each lattice and for each of \ref{itemI}--\ref{itemIII} we ran 20 independent threads. In each thread we collected samples after every 10 attempted BFACF moves, up to a total of $2\times10^{10}$ samples.

\subsection{Enumeration}\label{ssec:enumeration}

Our first goal is to estimate the number $\theta_n$ of thetas of size $n$ on each of the lattices in question. This can be achieved using any of \ref{itemI}--\ref{itemIII} above, summing over $m$, $s$ or $r$ respectively. Each thus gives a different estimate of $\theta_n$.

The known terms of the sequences $(\theta_n)$ are (to the best of our knowledge) 
\begin{equation}
 \begin{split} (\theta_n)_{n \geq 7} = (2, 0, 12, 6, 62, 60, 338, 430, 1966, 2794, 11772, 17898, 71390, 114496, 438112, \\ 731698, 2718114, 4681116, 17013354, 30025926, 107283688, 193174670, \dots) \end{split}
\end{equation}
for the square lattice, and
\begin{equation}
(\theta_n)_{n \geq 7} = (18, 24, 344, 582, 5934, 12120, 104250, 239610, 1877626, 4655982, \dots) 
\end{equation}
for the cubic lattice. These were computed using a basic backtracking algorithm.

Recall that with $p_n$ denoting the number of SAPs (of any knot type) of perimeter $n$ on a given lattice, it is widely expected that 
\begin{equation}\label{eqn:pn_asympts}
    p_n = C \mu^n n^{\alpha-3}(1+o(1)).
\end{equation}
Unknots are known \cite{sumners_knots_1988} to have a strictly smaller growth rate $\mu_0 < \mu$, and numerical evidence \cite{janse_van_rensburg_universality_2011} suggests that the exponent $\alpha_0$ is the same as for all polygons. For other fixed knot types $K$, it is conjectured \cite{janse_van_rensburg_universality_2011,baiesi_entropic_2010} that $\mu_K = \mu_0$ and $\alpha_K = \alpha_0 + f_K$, where $f_K$ is the number of prime knot components of $K$.
For knots in very narrow tubes of the cubic lattice, this result has been proved \cite{beaton_entanglement_2026}.

In two dimensions it is believed that the critical exponent $\alpha=\frac12$, while in three dimensions $\alpha \approx 0.23721$ \cite{guttmann_polygons_2009}. (There is no reason to believe the latter value is rational, or even algebraic.) The value $\mu$ is referred to as the \emph{connective constant} (or sometimes \emph{growth constant}). For the square and cubic lattices, the current best estimates are
\begin{equation}
    \mu \approx \begin{cases}
        2.63815853032790(3), & \text{square \cite{jacobsen_growth_2016}} \\
        4.684039931(27), & \text{cubic \cite{clisby_calculation_2013}} \\
    \end{cases}
\end{equation}
where the values in brackets indicate uncertainty in the final digit. 

As mentioned above, in this work we are considering ``unknotted'' thetas in the cubic lattice. As a result it is more appropriate to use the growth rate $\mu_0$ of unknots instead of $\mu$. Numerical estimates indicate that $\log\mu-\log\mu_0 \approx 4.15 \times 10^{-6}$ \cite{janse_van_rensburg_knot_1990,janse_van_rensburg_thoughts_2008}.
\footnote{Though for the lengths of thetas that we are working with, this difference has essentially no effect on any of our calculations.}

The lower order terms are expected  to have correction-to-scaling exponents:
\begin{equation}
    p_n = C \mu^n n^{\alpha-3}\left(1 + \frac{a_1}{n} + \frac{a_2}{n^2} + \cdots + \frac{b_0}{n^{\Delta_1}} + \frac{b_1}{n^{\Delta_1+1}} + \cdots \right)
\end{equation}
for a universal exponent $\Delta_1$. Numerical evidence \cite{guttmann_polygons_2009,caracciolo_correction--scaling_2005} suggests that $\Delta_1 = \frac32$ in two dimensions. We are not aware of numerical estimates for $\Delta_1$ using three dimensional self-avoiding \emph{polygons}, but work on self-avoiding \emph{walks} \cite{clisby_high-precision_2016} has estimated $\Delta_1 \approx 0.528(8)$.

For thetas it is known \cite{guttmann_two-dimensional_1978} that $\theta_n$ has the same exponential growth rate $\mu$. (Technically this result is for 2D lattices; however it can easily be generalised to 3D.) We will assume that $\theta_n$ has a similar subexponential factor form
\begin{equation}\label{eqn:thetan_asympts}
    \theta_n \sim B\mu^n n^\zeta.
\end{equation}
for constants $B$ and $\zeta$. (In this work we have made no attempt to estimate $B$.)

There is little work in the literature on the number of lattice thetas. They appear in Sykes' ``counting theorem'' \cite{sykes_counting_1961}, which relates the numbers of self-avoiding walks, polygons, thetas, tadpoles and figure-eights. Some numerical work was done in \cite{guttmann_two-dimensional_1978} on the square and triangular lattices, resulting in the estimate $\zeta = -1.35 \pm 0.15$ (they use the symbol $\delta$, corresponding to $\zeta=\delta-1$). Enumeration of thetas on the hexagonal lattice was used in \cite{guttmann_self-avoiding_2004} but asymptotics were not computed at the time. Some short series have also been provided to us \cite{guttmann_private_comm}.

As will be seen in \cref{ssec:polydisperse_armlengths}, a typical theta-graph of size $n$ tends to have two short arms of size $o(n)$ and one long arm of size $\approx n$. Roughly speaking, it follows that a typical theta resembles a SAP of size $\approx n$ with a small loop of size $o(n)$  ``inserted'' somewhere. Hence one might expect that $\theta_n \approx cnp_n$ for constant $c = \frac{B}{C}$. Combining \eqref{eqn:pn_asympts} and \eqref{eqn:thetan_asympts} would then give $\zeta = \alpha-2$, i.e.\ $\zeta = -\frac32$ in 2D and $\zeta \approx -1.76279$ in 3D. We note that the value $-\frac32$ is just within the range suggested in \cite{guttmann_two-dimensional_1978}.

To analyse the data we will assume the basic asymptotic form \eqref{eqn:thetan_asympts} holds, or potentially  one with the correction-to-scaling factor
\begin{equation}\label{eqn:theta_with_CtS}
    \theta_n \sim B\mu^n n^\zeta\left(1+\frac{a}{n^\Delta}\right)
\end{equation}
for a constant $a$, where we take $\Delta=1$ for 2D and $\Delta=\frac12$ for 3D. In fact, because the Wang-Landau method really produces estimates for $\log \theta_n$, we will actually use the $\log$s of \eqref{eqn:thetan_asympts} and \eqref{eqn:theta_with_CtS}
\begin{equation}\label{eqn:ln_with_cts}
    L_n = \log\theta_n \sim \log B + n\log\mu + \zeta\log n + \frac{a}{n^\Delta}
\end{equation}
where we have used $\log(1+\frac{a}{n^\Delta}) \sim \frac{a}{n^\Delta}$.

\subsubsection*{Square lattice}

For the square and cubic lattices we essentially have two sequences -- one for even $n$ and one for odd $n$. We expect $\mu$ and $\zeta$ to be the same for both, but $B$ and $a$ may depend on the parity of $n$. We can analyse these separately, but it is also fruitful to analyse the median sequence
\begin{align}
    L^*_n &:= \frac12 L_n + \frac12 L_{n+1} \\ 
    & \textstyle \sim \log B^* + n\log\mu + \zeta \log n + \frac{a^*}{n^\Delta}
\end{align}
where $B^*$ and $a^*$ are constants which depend on $B_\text{even}, B_\text{odd}, a_\text{even}, a_\text{odd}$, and $\mu$.

\begin{figure}
    \centering
    \begin{subfigure}{0.45\textwidth}
    \centering
    \includegraphics[width=\textwidth]{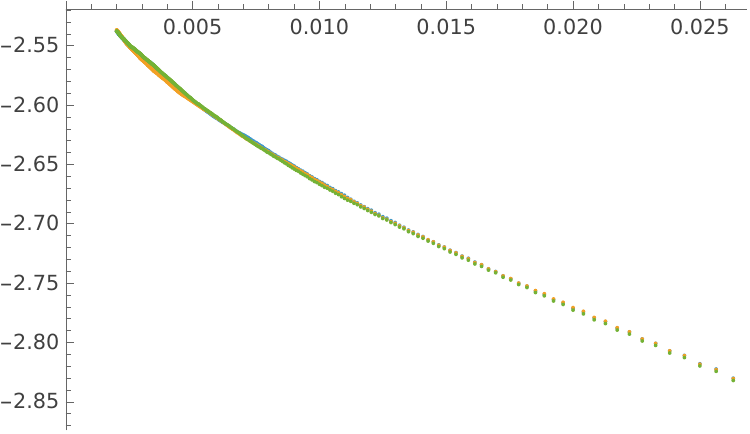}
    \caption{}
    \end{subfigure}
    \hfill
    \begin{subfigure}{0.45\textwidth}
    \centering
    \includegraphics[width=\textwidth]{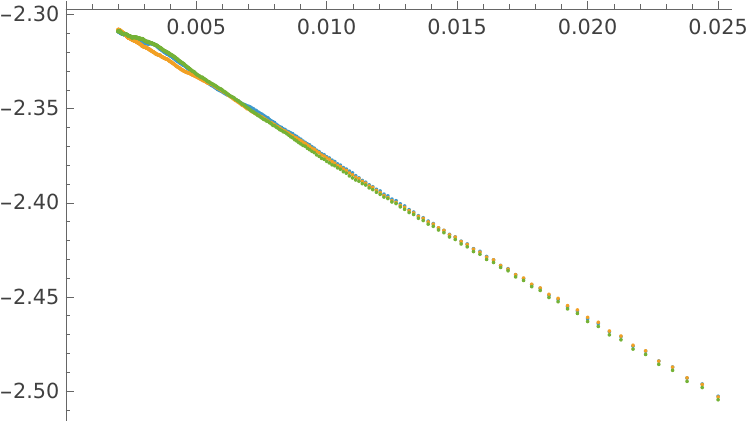}
    \caption{}
    \end{subfigure}
    \caption{\textbf{(a)} A plot of $L^*_n - n\log\mu - (\alpha-2) \log n$  for the square lattice against $\frac1n$. The data are from \ref{itemI} (blue), \ref{itemII} (orange) and \ref{itemIII} (green), taken by first averaging over the 20 independent Wang-Landau runs. \textbf{(b)} The same data, except with the value $-1.459$ used instead of $(\alpha-2)$.}
    \label{fig:sq_zeta_testing_different_values}
\end{figure}

We first test the proposition that $\zeta_\text{sq} = \alpha_\text{sq}-2 = -\frac32$ by plotting $L^*_n - n\log\mu - (\alpha-2) \log n$ against $\frac{1}{n}$. If indeed $\zeta=\alpha-2$ then (assuming $\Delta=1$) this plot should look linear. In \cref{fig:sq_zeta_testing_different_values} (a) this quantity is plotted (using data from \ref{itemI}--\ref{itemIII}, averaged over the independent Wang-Landau runs) and it is clearly not linear in $\frac1n$. On the other hand, by testing different values of $\zeta$ and minimising the sum of the residuals between the data and a linear fit, we find $\zeta_\text{sq} \approx -1.459$ results in a quite straight plot (see \cref{fig:sq_zeta_testing_different_values} (b)).

\begin{figure}
    \centering
    \begin{subfigure}{0.45\textwidth}
    \centering
    \includegraphics[width=\textwidth]{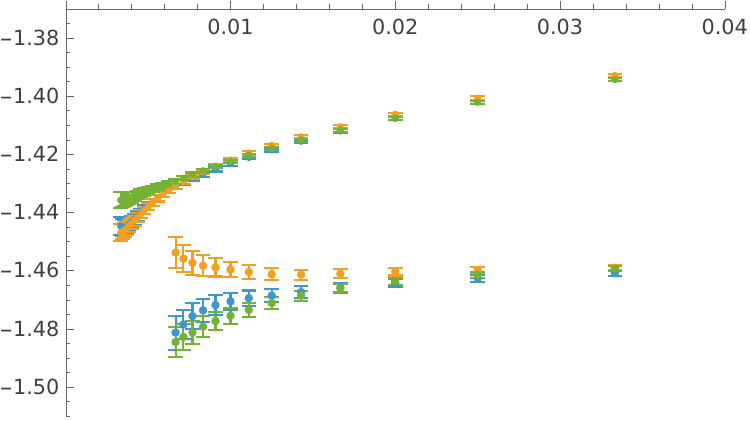}
    \caption{}
    \end{subfigure}
    \hfill
    \begin{subfigure}{0.45\textwidth}
    \centering
    \includegraphics[width=\textwidth]{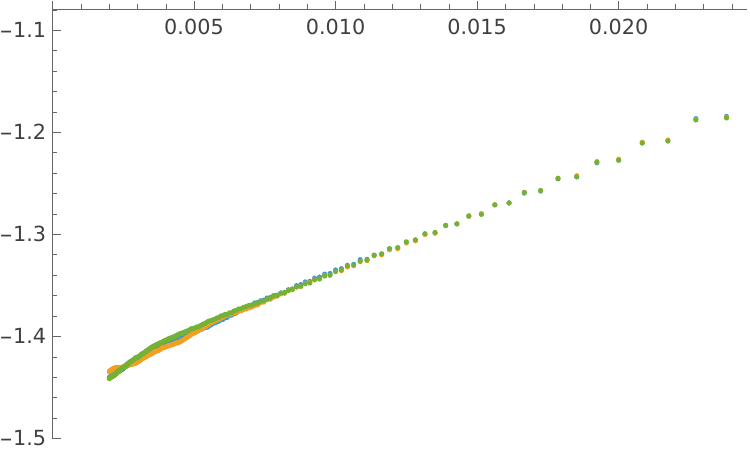}
    \caption{}
    \end{subfigure}
    \caption{\textbf{(a)} Plots of the estimated values of $\zeta$ for the square lattice, using median data from \ref{itemI} (blue), \ref{itemII} (orange) and \ref{itemIII} (green). These are obtained using \textsc{Mathematica}'s \textsc{LinearModelFit} function, fitting $L^*_n -n\log \mu$. The top three sets are fit without any correction-to-scaling term, while the bottom three do include the $\frac{a}{n^\Delta}$ term, with $\Delta=1$. The horizontal axis is $\frac{1}{\nmin}$.
    \textbf{(b)} A plot of $R^*_n$ as per \eqref{eqn:zeta_est_ratios}, for $n\in[20,500]$, using data from \ref{itemI} (blue), \ref{itemII} (orange) and \ref{itemIII} (green) computed by first averaging over the 20 independent Wang-Landau runs. The horizontal axis is $\frac1n$. A linear fit to these data has an intercept of $-1.458$.}
    \label{fig:sq_zeta_est_plots}
\end{figure}

We also tried directly fitting curves of the form $\log B^* + \zeta \log n$ and $\log B^* + \zeta \log n + \frac{a^*}{n}$ to $L^*_n - n\log \mu$, with mixed results. We separately took the data generated by \ref{itemI}--\ref{itemIII} and fitted curves to these using \textsc{Mathematica}'s \textsc{LinearModelFit} function (with the default 95\% confidence intervals) for values of $n$ in $[\nmin, 499]$. We varied $\nmin$ over the range $[10,300]$ for the basic asymptotic form and $[10,150]$ using the correction-to-scaling term. See \cref{fig:sq_zeta_est_plots} (a) for plots of the estimated values of $\zeta$, plotted against $\frac{1}{\nmin}$. Unfortunately these fits have not yielded particularly precise estimates.

Another method for estimating $\zeta$ is to note that
\begin{equation}\label{eqn:zeta_est_ratios}
    R^*_n := \frac{1}{\log 2}\left(L^*_n - L^*_{n/2} - \frac{n}{2}\log\mu\right) \sim  \zeta + \frac{a^\dagger}{n^\Delta}
\end{equation}
for a constant $a^\dagger$ which depends on $a^*$ and $\Delta$. (Note that we need to use $L^*_n$ here instead of $L_n$, otherwise the difference between $B_\text{even}$ and $B_\text{odd}$ introduces a further term.) In \cref{fig:sq_zeta_est_plots} (b) we plot $R^*_n$ for $n\in[20,500]$ using data from \ref{itemI}--\ref{itemIII}, computed by first averaging over the 20 independent Wang-Landau runs. The three sets of data are quite linear in $\frac1n$, as expected. By taking a linear fit to these data, the projected vertical intercept is $-1.458$.

Based on the data plotted in \cref{fig:sq_zeta_testing_different_values} (b) and \cref{fig:sq_zeta_est_plots} (b),  we estimate that $\zeta_\text{sq} = -1.458 \pm 0.005$. It seems quite likely to us that $\zeta_\text{sq} > -\frac32$, that is, $\zeta_\text{sq} > \alpha-2$ and our rough calculation earlier in this section was missing some details. It is, of course, possible that with samples of much larger thetas, estimates of $\zeta$ may yet get closer to $\alpha-2$. We also note that virtually every exponent related to 2D SAPs, SAWs, etc., is a rational number whose denominator is a power of 2; however, at this point we do not have a sufficiently precise estimate for $\zeta_\text{sq}$ to conjecture such a value.

To shed some further light on the possibility that $\zeta_\text{sq} > \alpha-2$, we can attempt to investigate the behaviour of the number of thetas whose shortest arm (or second-shortest, or both) is as small as possible. These may be more likely to look like a large SAP with a very small loop inserted somewhere. 

For a given theta-graph $T$, let $(\ell_1(T),\ell_2(T),\ell_3(T))$ be the number of edges in the three arms, with $\ell_1 \leq \ell_2 \leq \ell_3$. Then, we define the counting sequences for thetas with $\ell_1$, $\ell_2$ or $\ell_1+\ell_2$ minimal:
\begin{align}
    \ult^{[1]}_n & = |\{T \in \mathcal{T}_n : \ell_1(T) = \textstyle \frac12(3+(-1)^n)\}| \\
    \ult^{[2]}_n &= |\{T \in \mathcal{T}_n : \ell_2(T) = \textstyle \frac12(5-(-1)^n)\}| \\
    \ult^{[12]}_n &= |\{T \in \mathcal{T}_n : \ell_1(T) + \ell_2(T) = 4\}|.
\end{align}
(These definitions work for both the square and cubic lattices.) Then we expect
\begin{equation}
    \ult^{[1]}_n \sim C \mu^n n^\lambda
\end{equation}
for an exponent $\lambda$, which is likely to be $\alpha-2$. Similar behaviour should hold for $\theta^{[2]}_n$ and $\theta^{[12]}_n$. As before we also use the $*$ superscript to denote the median sequences.

\begin{figure}
    \centering
    \begin{subfigure}{0.45\textwidth}
    \centering
    \includegraphics[width=\textwidth]{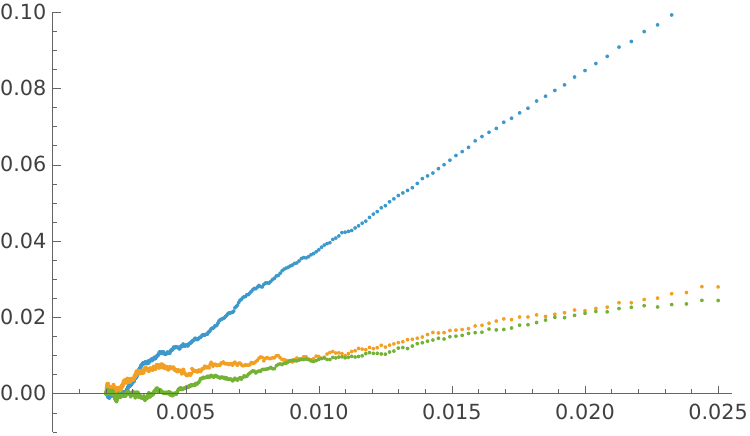}
    \caption{}
    \end{subfigure}
    \hfill
    \begin{subfigure}{0.45\textwidth}
    \centering
    \includegraphics[width=\textwidth]{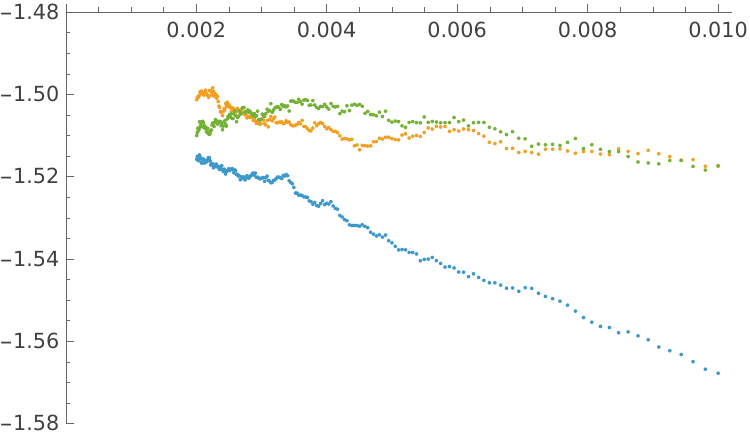}
    \caption{}
    \end{subfigure}
    \caption{\textbf{(a)} Plots of $\ult^{[1]*}_n - n\log\mu - (\alpha-2)$ (blue) and similarly for $\ult^{[2]*}_n$ (orange) and $\ult^{[12]*}_n$ (green), all against $\frac1n$, for the square lattice. These have been shifted vertically so that the final term is 0 (so that they all fit on the same plot). The plots are linear as expected. \textbf{(b)} A similar plot to \cref{fig:sq_zeta_est_plots} (b), using ratios to estimate $\lambda$. (The colour scheme is the same as \textbf{(a)}.) Using linear fits, the projected intercepts are $-1.501$ (blue), $-1.502$ (orange) and $-1.499$ (green).}
    \label{fig:sq_lambda_est_plots}
\end{figure}

In \cref{fig:sq_lambda_est_plots} (a) we plot $\ult^{[1]*}_n - n\log\mu - (\alpha-2)$ (blue) and similarly for $\ult^{[2]*}_n$ (orange) and $\ult^{[12]*}_n$ (green). As expected, the data are quite linear in $\frac1n$, suggesting that $\alpha-2$ is the correct exponent. In \cref{fig:sq_lambda_est_plots} (b) we plot the analogous quantity to \eqref{eqn:zeta_est_ratios}.  While the plots are not as clean as those in \cref{fig:sq_zeta_est_plots} (b), we think it very likely that $\lambda = \alpha-2=-\frac32$. So this contrasts with our estimate of $\zeta_\text{sq}$, further suggesting that the argument that $\zeta$ is $\alpha-2$ is an oversimplification.

\subsubsection*{Cubic lattice}

\begin{figure}
    \centering
    \begin{subfigure}{0.45\textwidth}
    \centering
    \includegraphics[width=\textwidth]{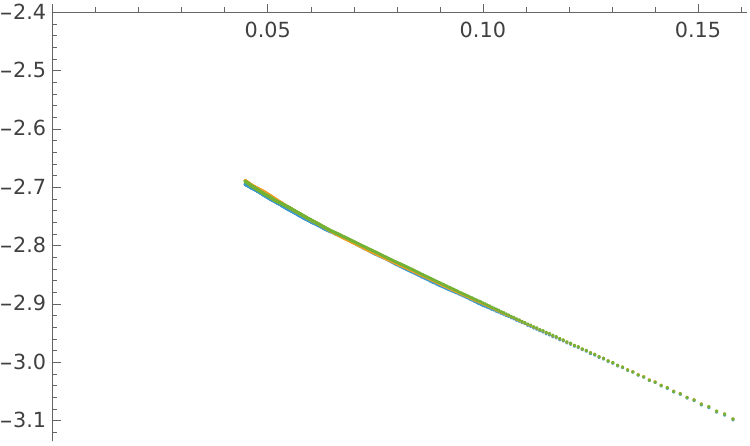}
    \caption{}
    \end{subfigure}
    \hfill
    \begin{subfigure}{0.45\textwidth}
    \centering
    \includegraphics[width=\textwidth]{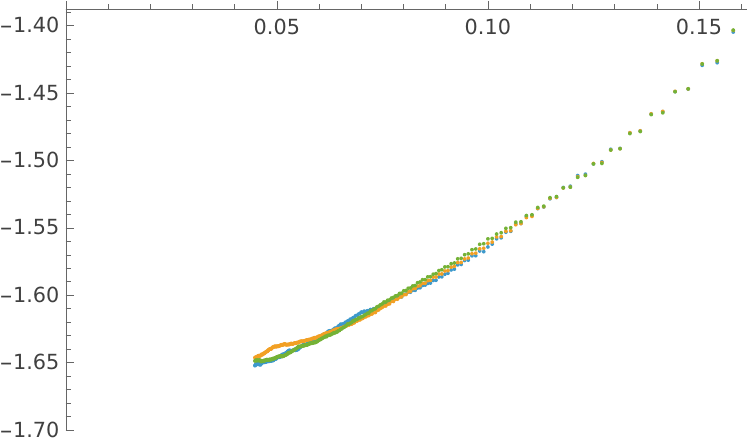}
    \caption{}
    \end{subfigure}
    \caption{\textbf{(a)} A plot of $L^*_n - n\log\mu_0 - (\alpha-2) \log n$  for the cubic lattice against $\frac{1}{\sqrt{n}}$. The data are from \ref{itemI} (blue), \ref{itemII} (orange) and \ref{itemIII} (green), taken by first averaging over the 20 independent Wang-Landau runs. \textbf{(b)} A plot of $R^*_n$ as per \eqref{eqn:zeta_est_ratios}, using data from \ref{itemI} (blue), \ref{itemII} (orange) and \ref{itemIII} (green) computed by first averaging over the 20 independent Wang-Landau runs. The horizontal axis is $\frac{1}{\sqrt{n}}$.}
    \label{fig:cubic_zeta_est_plots_using_sqrtn}
\end{figure}

For the cubic lattice we repeat the calculations from above. We initially work with the assumption that $\Delta \approx \frac12$. In \cref{fig:cubic_zeta_est_plots_using_sqrtn} we plot similar quantities to \cref{fig:sq_zeta_testing_different_values} (a) and \cref{fig:sq_zeta_est_plots} (b), plotting against $\frac{1}{\sqrt{n}}$. The plot in \cref{fig:cubic_zeta_est_plots_using_sqrtn} (a) displays some curvature; changing the exponent from $\alpha-2$ to something larger did not yield a nicely straight curve. However, \cref{fig:cubic_zeta_est_plots_using_sqrtn} (b) is more telling -- it is decidedly not straight, indicating that $\frac{1}{\sqrt{n}}$ is not the correct correction-to-scaling form.

\begin{figure}
    \centering
    \begin{subfigure}{0.45\textwidth}
    \centering
    \includegraphics[width=\textwidth]{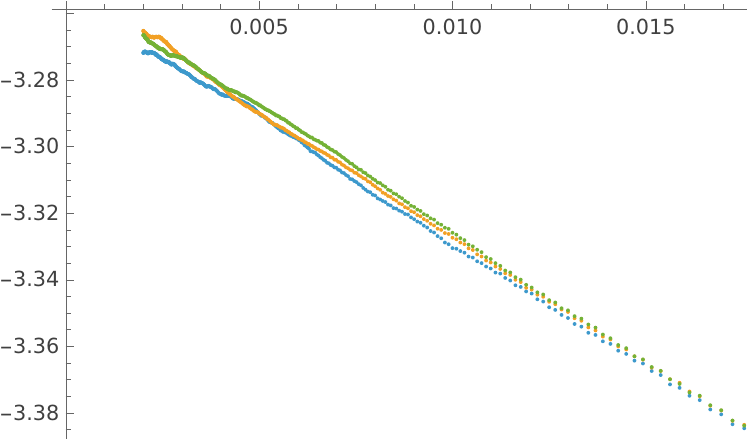}
    \caption{}
    \end{subfigure}
    \hfill
    \begin{subfigure}{0.45\textwidth}
    \centering
    \includegraphics[width=\textwidth]{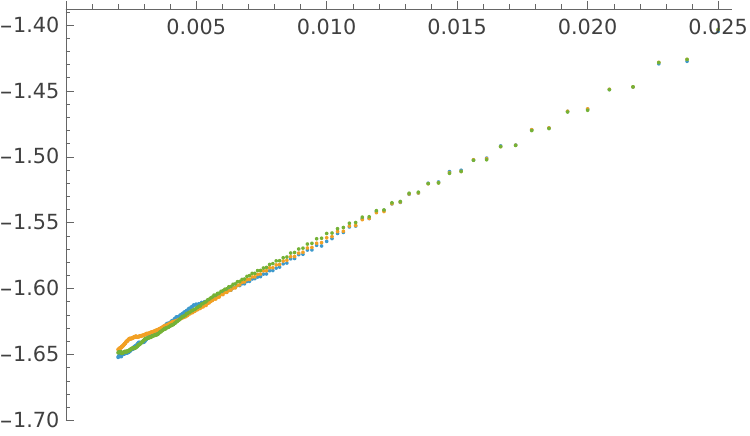}
    \caption{}
    \end{subfigure}
    \caption{\textbf{(a)} A plot of $L^*_n - n\log\mu_0 - (-1.67) \log n$  for the cubic lattice against $\frac{1}{n}$. The data are from \ref{itemI} (blue), \ref{itemII} (orange) and \ref{itemIII} (green), taken by first averaging over the 20 independent Wang-Landau runs. \textbf{(b)} A plot of $R^*_n$ as per \eqref{eqn:zeta_est_ratios}, using data from \ref{itemI} (blue), \ref{itemII} (orange) and \ref{itemIII} (green) computed by first averaging over the 20 independent Wang-Landau runs. The horizontal axis is $\frac{1}{n}$. A linear fit to these data has intercept $-1.669$.}
    \label{fig:cubic_zeta_est_plots_using_1n}
\end{figure}

As a result of this uncertainty we instead tried using the correction-to-scaling exponent $\Delta=1$. The results are plotted in \cref{fig:cubic_zeta_est_plots_using_1n}. For (a) we tried different values of $\zeta$ in order to minimise the sum of the residuals between the data and a linear fit, and settled on $-1.670$ as the best value. In (b) we plot $R_n^*$ against $\frac1n$, and found a quite straight plot. A linear fit to these data has intercept $-1.669$.

Based on the plots in \cref{fig:cubic_zeta_est_plots_using_sqrtn,fig:cubic_zeta_est_plots_using_1n}, we think it likely that the correction-to-scaling exponent $\Delta=1$ is more appropriate for estimating $\zeta$ for cubic lattice thetas. With this value, the data supports an estimate of the entropic exponent $\zeta_\text{cu} = -1.67\pm 0.01$. This is quite different to $\alpha-2 = -1.763.$

We also looked at the behaviour of $\ult_n^{[1]*}$, $\ult_n^{[2]*}$ and $\ult_n^{[12]*}$ for the cubic lattice. We have omitted the plots for brevity. We again found that plotting against $\frac1n$ yielded straighter plots than $\frac{1}{\sqrt{n}}$. With the assumption $\Delta=1$ then the exponent $\lambda$ does indeed appear to be close to $\alpha-2$.

\begin{rem}
    Define $\theta_{\ell_1,\ell_2,n-\ell_1-\ell_2}$ to be the number of (unknotted) thetas with arm lengths $\ell_1$, $\ell_2$ and $n-\ell_1-\ell_2$, ordered according to length (assuming the three lengths have the same parity). After conversation with Stuart Whittington \cite{whittington_private_comm}, we expect that a pattern theorem can be used to show that, for fixed $\ell_1$ and $\ell_2$ and as $n\to\infty$, there exist positive constants $A,B$ such that
    \begin{equation}
        Anp_n \leq \theta_{\ell_1,\ell_2,n-\ell_1-\ell_2} \leq Bnp_n
    \end{equation}
    for even $n$, with a similar result holding for odd $n$.
    This implies that the critical exponent for thetas with the two shortest arms of fixed length is indeed equal to $\alpha-2$. This argument can be applied to both the square and cubic lattices.
\end{rem}

\subsection{Distribution of arm-lengths}\label{ssec:polydisperse_armlengths}

We next look at the distribution of arm-lengths in polydisperse thetas.  Let $\langle \ell_1 \rangle_n$ be the average number of edges in the shortest arm of thetas of total size $n$. We can similarly consider $\langle \ell_2 \rangle_n$ and $\langle \ell_1+\ell_2\rangle_n = n - \langle \ell_3 \rangle_n$. We will use $\langle \ell_{12}\rangle_n$ to denote this latter value.

For $\ell_1$ we use data generated in \ref{itemI}. It is not unreasonable to expect that
\begin{equation}\label{eqn:sigma_scaling_generic}
    \langle \ell_1 \rangle_n \sim C n^\sigma
\end{equation}
for some constants $C,\sigma$, where $\sigma$ is (possibly) universal and depends only on dimension. In 2D we assume a correction term of the form
\begin{equation}\label{eqn:sigma_scaling_with_correction}
    \langle \ell_1 \rangle_n \sim C n^\sigma \left(1 + \frac{s}{n}\right) \qquad \Rightarrow \qquad \log \langle \ell_1 \rangle_n \sim \log C + \sigma\log n + \frac{s}{n}
\end{equation}
for a constant $s$. For 3D we consider both $\frac1n$ and $\frac{1}{\sqrt{n}}$ as possible correction terms. As with the scaling of the number of thetas, for the square and cubic lattices it is possible that the constants $C$ and $r$ depend on the parity of $n$. We will thus make use of the median sequence $\langle \ell_1 \rangle_n^* = \frac12 \left(\langle \ell_1 \rangle_n + \langle \ell_1 \rangle_{n+1}\right)$.

For $\langle \ell_2 \rangle_n$ we similarly anticipate
\begin{equation}\label{eqn:tau_scaling_with_correction}
    \langle \ell_2 \rangle_n \sim D n^\tau
\end{equation}
for constants $D, \tau$. For numerical estimates we assume the same generic $\frac1n$ correction-to-scaling term. For $\langle \ell_{12} \rangle_n$, the exponent $\tau$ (if it exists) must be the same as for $\langle \ell_2 \rangle_n$, but the constant $D$ may differ. 

For the numerical analysis (estimating $\sigma$ and $\tau$) we performed essentially the same calculations as in \cref{ssec:enumeration} when estimating $\zeta$. We fit expressions of the form \eqref{eqn:sigma_scaling_with_correction}
or an expression similar to \eqref{eqn:zeta_est_ratios}:
\begin{equation}\label{eqn:ratios_sigma}
    \frac{1}{\log 2}\left(\log\langle \ell_1 \rangle_n^* - \log\langle \ell_1 \rangle_{n/2}^*\right) \sim \sigma + \frac{s^\dagger}{n}
\end{equation}
for a constant $s^\dagger$.

\subsubsection*{Square lattice}

\begin{figure}
    \centering
    \begin{subfigure}{0.45\textwidth}
    \centering
    \includegraphics[width=\textwidth]{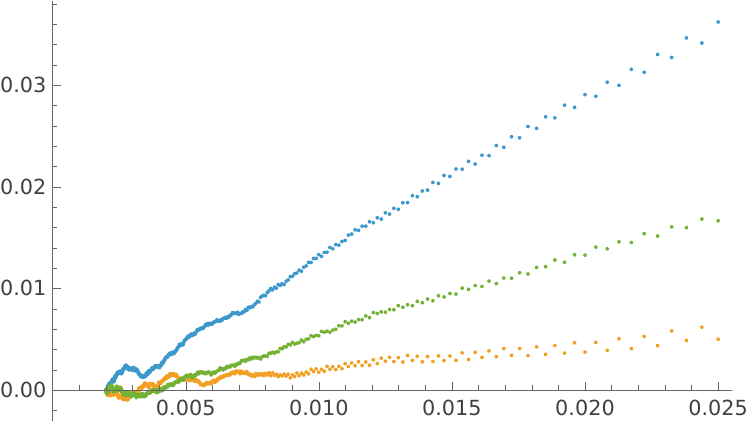}
    \caption{}
    \end{subfigure}
    \hfill
    \begin{subfigure}{0.45\textwidth}
    \centering
    \includegraphics[width=\textwidth]{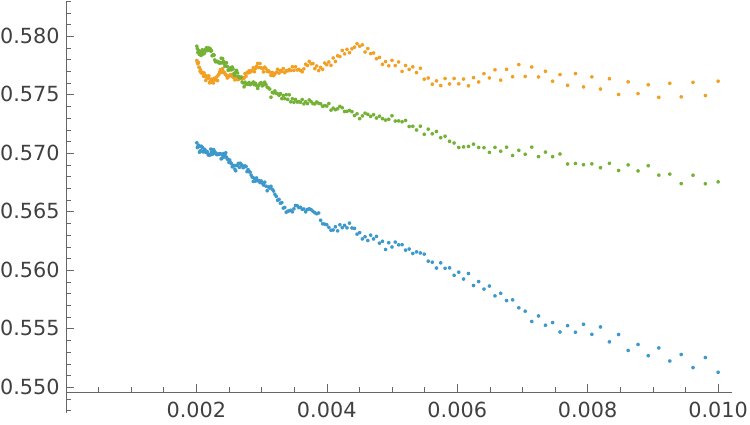}
    \caption{}
    \end{subfigure}
    \caption{\textbf{(a)} Plots of $\langle \ell_1\rangle_n^* - 0.574\log n$ (blue), $\langle \ell_2\rangle_n^* - 0.579\log n$ (orange) and $\langle \ell_{12}\rangle_n^* - 0.579\log n$ (green) against $\frac1n$ for the square lattice. These have been shifted vertically so that the final term is 0. \textbf{(b)} Plots of the analogous quantities to \eqref{eqn:ratios_sigma}, with the same colour schemes. Linear fits have intercepts 0.575 (blue), 0.577 (orange) and 0.581 (green).} 
    \label{fig:sq_sigma_tau_est_plots}
\end{figure}

For the square lattice we use the same approach as in the previous section, this time to estimate $\sigma$ and $\tau$. The results are plotted in \cref{fig:sq_sigma_tau_est_plots}. By plotting $\langle \ell_1\rangle^*_n - \sigma \log n$ against $\frac1n$ for different values of $\sigma$, and minimising the total residuals between these data and a line of best fit, we arrive at $\sigma = 0.574$. (See \cref{fig:sq_sigma_tau_est_plots} (a).) For $\tau$ we do the same with $\langle \ell_2\rangle^*_n$ and $\langle \ell_{12}\rangle^*_n$ (we fit both separately using the same $\tau$ value, and added the residuals for both). This gave the best estimate $\tau=0.579$. In \cref{fig:sq_sigma_tau_est_plots} (b) we plot the ratio quantity \eqref{eqn:ratios_sigma} and similarly for $\ell_2$ and $\ell_{12}$. 

From all these plots we think it more likely than not that $\sigma_\text{sq}=\tau_\text{sq}$, with a value of about $0.577 \pm 0.005$. We do note that $\frac{37}{64} \approx 0.578125$ is definitely in the vicinity of our estimate.

In \cref{fig:sq_cubic_shortest_distribution} (a) we plot the distribution of $\ell_1$ (i.e.\ the fraction of size $n$ thetas with a given value of $\ell_1$) for a range of $n$. Note the vertical scale -- the numbers drop off rapidly as $\ell_1$ increases.

\subsubsection*{Cubic lattice}

\begin{figure}
    \centering
    \begin{subfigure}{0.45\textwidth}
    \centering
    \includegraphics[width=\textwidth]{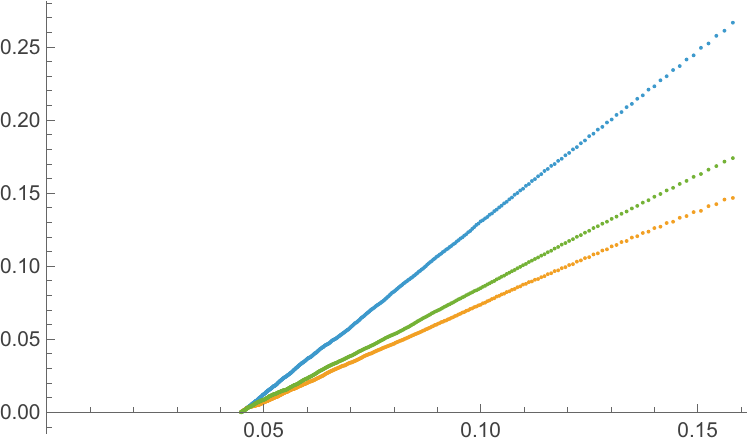}
    \caption{}
    \end{subfigure}
    \hfill
    \begin{subfigure}{0.45\textwidth}
    \centering
    \includegraphics[width=\textwidth]{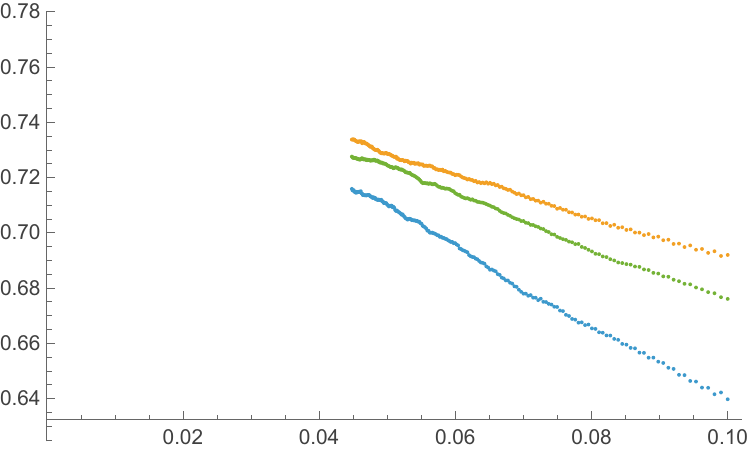}
    \caption{}
    \end{subfigure}
    \caption{\textbf{(a)} Plots of $\langle \ell_1\rangle_n^* - 0.779\log n$ (blue), $\langle \ell_2\rangle_n^* - 0.769\log n$ (orange) and $\langle \ell_{12}\rangle_n^* - 0.769\log n$ (green) against $\frac{1}{\sqrt{n}}$ for the cubic lattice. These have been shifted vertically so that the final term is 0. \textbf{(b)} Plots of the analogous quantities to \eqref{eqn:ratios_sigma}, with the same colour schemes, plotted against $\frac{1}{\sqrt{n}}$. Linear fits have intercepts 0.781 (blue), 0.768 (orange) and 0.773 (green).} 
    \label{fig:cubic_sigma_tau_est_plots}
\end{figure}

For the cubic lattice we repeat the calculations above. This time using the correction exponent $\Delta=\frac12$ yields much straighter plots than $\Delta=1$. See \cref{fig:cubic_sigma_tau_est_plots}.

In \cref{fig:cubic_sigma_tau_est_plots} (a) we get the best straight fit when $\sigma=0.779$ and $\tau=0.769$. Of course it is nonsensical for $\sigma>\tau$, but we expect that this anomaly just comes down to numerical uncertainty and a small-$n$ effect. It seems likely that $\sigma_\text{cu} = \tau_\text{cu}$ and this value is in the region of $0.775 \pm 0.01$.

\begin{figure}
    \centering
    \begin{subfigure}{0.45\textwidth}
    \centering
    \includegraphics[width=\textwidth]{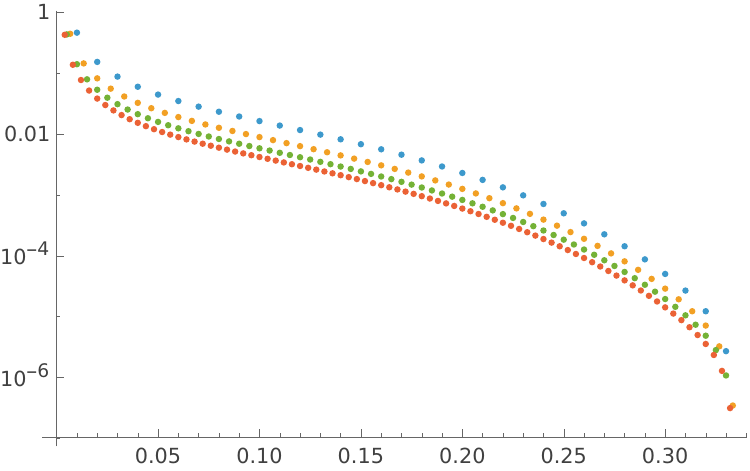}
    \caption{}
    \end{subfigure}
    \hfill
    \begin{subfigure}{0.45\textwidth}
    \centering
    \includegraphics[width=\textwidth]{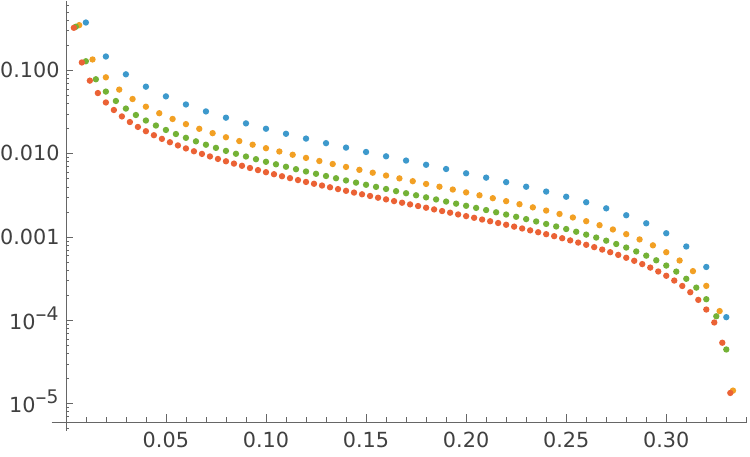}
    \caption{}
    \end{subfigure}
    \caption{\textbf{(a)} The distribution of $\ell_1$ (i.e.\ the fraction of thetas with a given $\ell_1$) across thetas of size $n=200$ (blue), 300 (orange), 400 (green) and 500 (red). The horizontal axis has been scaled by $n$. \textbf{(b)} The same plot for the cubic lattice.}
    \label{fig:sq_cubic_shortest_distribution}
\end{figure}

In \cref{fig:sq_cubic_shortest_distribution} (b) we plot the distribution of $\ell_1$ for thetas of various sizes on the cubic lattice. While the drop is not quite as precipitous as for the square lattice, cubic lattice thetas are still very much dominated by those with small $\ell_1$.

We note here that $\sigma \approx 0.775$ is not far off the estimated exponent $t=0.75$ for the average size of the knotted component of a prime knot, as reported in \cite{marcone_size_2007}. It is however a fair bit larger than the other estimates (0.4 to 0.65) as outlined in \cref{sec:introduction}. The distribution of $\ell_1$ in \cref{fig:sq_cubic_shortest_distribution} (b) can also be contrasted with the distribution of knot sizes in linear chains as found in \cite{tubiana_spontaneous_2013}, which showed a peak at knot-size around 100--300 (possibly independent of chain length), followed by a power-law decay. Here the distribution of $\ell_1$ is monotone decreasing, with the smallest values being the most populous.

\subsection{Separation of the branch points}

Various geometric quantities for self-avoiding walks and polygons, such as the squared radius of gyration and squared end-to-end distance, are expected to scale (in mean) as a power law with exponent $2\nu$. For example, the squared radius of gyration scales as
\begin{equation}
    \langle R_g^2 \rangle_n \sim Cn^{2\nu}\left(1+\frac{a}{n^\Delta}\right)
\end{equation}
for constants $C$ and $a$. The exponent $\nu$ is believed to be universal, taking values $\nu=\frac34$ in two dimensions and $\nu=0.587597(7)$ in three dimensions \cite{clisby_accurate_2010}.

Here we investigate the separation of the two vertices of degree 3. For a theta $T$ we write $D^2(T)$ to be the squared distance between the two vertices of degree 3. Then it is reasonable to expect that the mean of this quantity (across all thetas of size $n$) is
\begin{equation}\label{eqn:D2_asymp}
    \langle D^2\rangle_n \sim Cn^{2\eta}
\end{equation}
for some constants $C$ and $\eta$. It is our goal here to estimate $\eta$.

To facilitate this calculation, we took the approximate theta counts generated in \ref{itemI}--\ref{itemIII} and ran Markov chains with transition probabilities given by the Metropolis-Hastings algorithm (with Hastings factors as per \eqref{eqn:Hastings_factors}). For each of \ref{itemI}--\ref{itemIII} we ran 20 independent Markov chains, collecting $10^9$ samples in each, for the square and the cubic lattices. As in the previous sections there is dependence on the parity of $n$, so we also make use of the median sequence 
\begin{equation}
    \langle D^2\rangle_n^* = \frac12\left(\langle D^2\rangle_n +\langle D^2\rangle_{n+1}\right).
\end{equation}

\begin{figure}
    \centering
    \begin{subfigure}{0.45\textwidth}
    \centering
        \includegraphics[width=\textwidth]{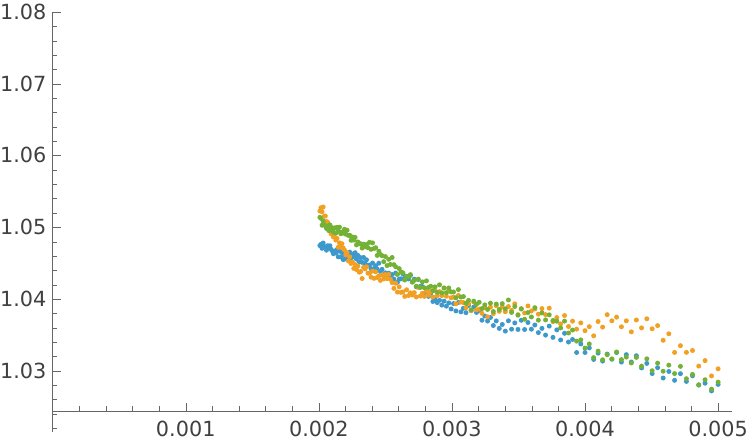}
    \end{subfigure}
    \hfill
    \begin{subfigure}{0.45\textwidth}
    \centering
        \includegraphics[width=\textwidth]{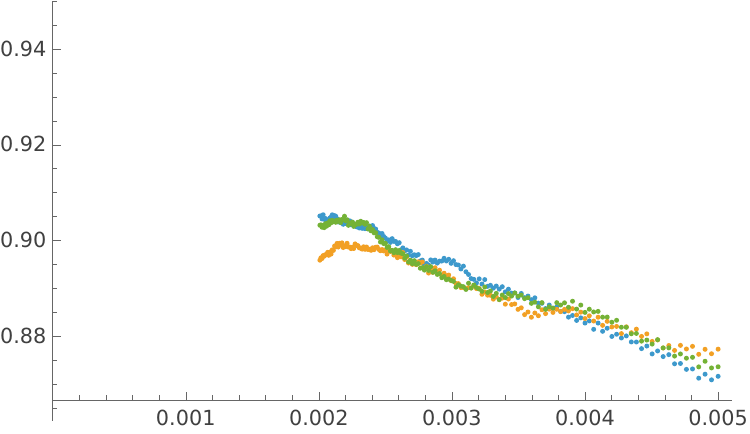}
    \end{subfigure}
    \caption{Ratio plots for estimating $2\eta$ on the \textbf{(a)} square lattice and \textbf{(b)} cubic lattice, both plotted against $\frac1n$ and using the median data in all cases. The Metropolis-Hastings transition probabilities were calculated using estimates from \ref{itemI} (blue), \ref{itemII} (orange) and \ref{itemIII} (green).}
    \label{fig:est_eta_ratios}
\end{figure}

We used several methods for estimating $2\eta$, including directly fitting to an expression of the form \eqref{eqn:D2_asymp} as well as using the same kind of ratio method as in \eqref{eqn:ratios_sigma}. See \cref{fig:est_eta_ratios} for ratio plots. For the square lattice we estimate $2\eta_\text{sq} = 1.06 \pm 0.01$. For the cubic lattice we have been unable to determine with much certainty whether $\Delta=\frac12$ or $\Delta=1$ is more appropriate for the correction-to-scaling term; both can be made to fit the data reasonably well. In \cref{fig:est_eta_ratios} we plot the ratios estimates against $\frac1n$; these give a value of about $2\eta_\text{cu} = 0.92 \pm 0.01$. If we instead use $\frac{1}{\sqrt{n}}$ (plot omitted for brevity) then the estimate increases somewhat, to around $2\eta_\text{cu} = 0.95 \pm 0.01$.

We note that the number of edges between the two vertices of degree 3 is typically $O(n^\sigma)$, where estimates for $\sigma$ were found in the previous section. Then
\begin{equation}
    \frac{\eta}{\sigma} \approx \begin{cases} 0.92, & \text{square} \\ [0.59,0.62], & \text{cubic.} \end{cases}
\end{equation}
These values should be contrasted with the values for $\nu$. They suggest that in two dimensions, the vertices of degree 3 are, on average, further apart than would be ``typical'' for two vertices separated by $O(n^\sigma)$ edges (say, in a SAW or a SAP). On the other hand, in three dimensions the value of $\frac{\eta}{\sigma}$ seems to be quite close to $\nu$, suggesting that there is little or no additional repulsion between the vertices of degree 3.

\section{Results: monodisperse theta-graphs}\label{sec:equilateral}

\begin{figure}
    \centering
    \begin{subfigure}{0.45\textwidth}
    \centering
        \includegraphics[width=\textwidth]{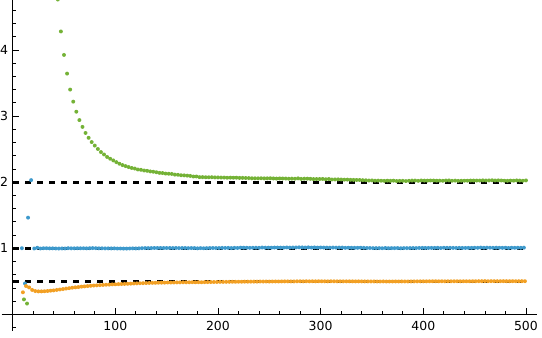}
    \end{subfigure}
    \hfill
    \begin{subfigure}{0.45\textwidth}
    \centering
        \includegraphics[width=\textwidth]{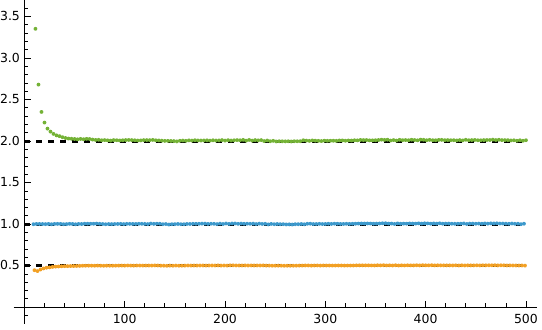}
    \end{subfigure}
    \caption{The ratio $\olt_n^{[12]}/\olt_n^{[1]}$ for the square lattice \textbf{(a)} and the cubic lattice \textbf{(b)}, with $n \equiv 0$ (mod 3) (blue), $n \equiv 1$ (mod 3) (orange), and $n \equiv 2$ (mod 3) (green). The ratios were computed by first averaging over the 10 independent Wang-Landau runs.}
    \label{fig:tn_sn_ratios}
\end{figure}

In this section we consider \emph{monodisperse} theta-graphs, where the three arms have the same length. Since this requires the total length to be a multiple of 3, we will relax the definition slightly, and define two related sequences. First define
\begin{align}
    M_1(n) &= \max_{T\in\mathcal{T}_n} \{\ell_1(T)\}, \\
    M_{12}(n) &= \max_{T\in\mathcal{T}_n} \{\ell_1(T)+\ell_2(T)\}.
\end{align}
It is not difficult to determine $M_1(n)$ and $M_{12}(n)$ for the square and cubic lattices. \nrb{could put these in an appendix?}
Then let
\begin{align}
    \olt_n^{[1]} & = |\{T \in \mathcal{T}_n : \ell_1(T) = M_1(n)\}| \\
    \olt_n^{[12]} &= |\{T \in \mathcal{T}_n : \ell_1(T)+\ell_2(T) = M_{12}(n)\}|
\end{align}
That is, $\olt_n^{[1]}$ counts those theta-graphs of size $n$ whose shortest arm is as long as possible, while $\olt_n^{[12]}$ counts those for which the sum of the two shortest arms is as large as possible (equivalently, the longest arm is as short as possible). The relationship between these depend on whether $ n$ (mod 3)  is equal to $0,1,2$. It is easy to show that
\begin{equation}
    \olt_n^{[1]} = \olt_n^{[12]} \qquad \text{if } n\equiv 0\text{ (mod 3) and } n\geq 21 \text{ (square lattice) or } n\geq 9 \text{ (cubic lattice)}.
\end{equation}
As an interesting aside, based on data from \ref{itemI} and \ref{itemII} we make the following conjecture, for which we have no combinatorial explanation. See \cref{fig:tn_sn_ratios}.
\begin{conj}\label{conj:tn_sn_ratio}
    For both the square and cubic lattices, as $n\to\infty$, 
    \begin{equation}
        \olt_n^{[12]} \sim \begin{cases} \frac12 \olt_n^{[1]} & \text{if } n \equiv 1 \text{ (mod 3)} \\
        2\olt_n^{[1]} & \text{if } n \equiv 2 \text{ (mod 3).} \end{cases}
    \end{equation}
\end{conj}

For monodisperse thetas we expect
\begin{equation}\label{eqn:sn_asymps}
    \olt_n^{[1]} \sim C \mu^n n^\beta
\end{equation}
for some exponent $\beta$, and where $C$ may depend on the value of $n$ (mod 3). In light of \cref{conj:tn_sn_ratio}, a similar expression should hold for $\olt_n^{[12]}$. It is our goal in this section to estimate $\beta$.

\subsubsection*{Square lattice}

Two-dimensional polymer networks with a fixed typology were studied by Duplantier in \cite{duplantier_polymer_1986} using renormalisation theory and conformal invariance. Thetas are a particular case of `watermelons' as considered there. For monodisperse thetas, the exponent $\beta$ as in \eqref{eqn:sn_asymps} corresponds to setting $L=3$ in \cite[Eqn. 23]{duplantier_polymer_1986}:
\begin{equation}
    \frac{20-9L^2}{32} - (L-1) = -\frac{125}{32} \approx -3.90625.
\end{equation}

\begin{figure}
    \centering
    \begin{subfigure}{0.45\textwidth}
    \centering
        \includegraphics[width=\textwidth]{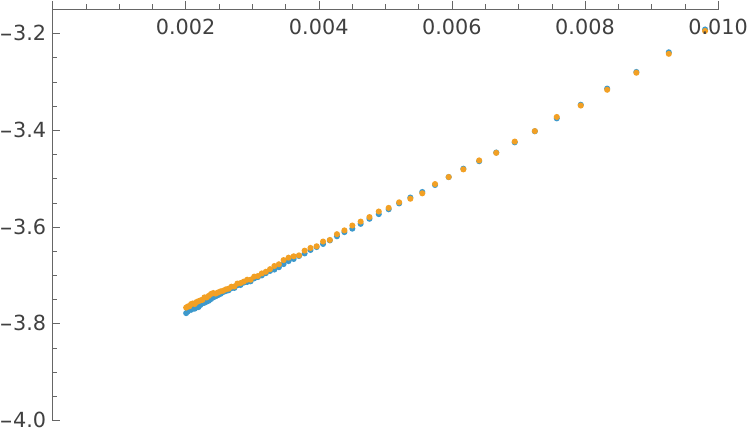}
    \end{subfigure}
    \hfill
    \begin{subfigure}{0.45\textwidth}
    \centering
        \includegraphics[width=\textwidth]{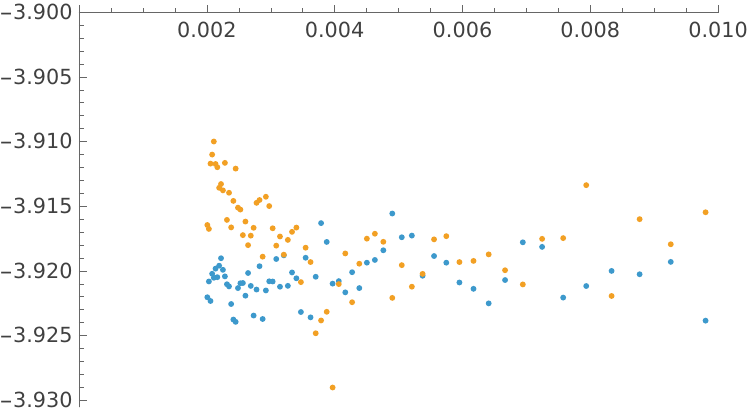}
    \end{subfigure}
    \caption{\textbf{(a)} A ratio plot for $\olt_n^{[1]}$ (blue) and $\olt_n^{[12]}$ (orange) on the square lattice, plotted against $\frac1n$ (using only values of $n$ which are multiples of 6). Taking linear fits through the last 50 points of both gives the intercept $-3.908$. \textbf{(b)} The equivalent plot for the cubic lattice.}
    \label{fig:est_beta_ratios}
\end{figure}

We have performed similar analyses as in the previous sections for estimating $\beta$. Plotting $\olt_n^{[1]} - n\log \mu - \beta\log n$ against $\frac1n$ for different values of $\beta$ (note that because we are only working with values of $n$ which are multiples of 3, we do not use the median data here), and likewise for $\olt_n^{[12]}$, and then finding the value which gives the most linear data (plot omitted for brevity) gives $\beta=-3.912$ as the optimal value. Using the ratio technique (see \cref{fig:est_beta_ratios} (a)) gives plots which are not quite linear for small $n$, but trend towards linearity for large $n$. Taking the last 50 values and extrapolating linear fits gives the intercept $-3.908$.

We thus estimate $\beta_\text{sq} = -3.910 \pm 0.005$, and so this strongly supports the prediction of Duplantier \cite{duplantier_polymer_1986} that $\beta_\text{sq} = -\frac{125}{32}$.

\subsubsection*{Cubic lattice}

For the cubic lattice we attempted the same analysis, see \cref{fig:est_beta_ratios} (b). We are unable to determine whether $\Delta=\frac12$ or $\Delta=1$ is the more appropriate correction-to-scaling exponent here. Our best estimate is $\beta_\text{cu} = -3.915 \pm 0.01$. (We have attempted other methods for estimating $\beta_\text{cu}$, resulting in similar estimates.)  It is intriguingly not impossible that $\beta_\text{cu} = \beta_\text{sq}$ and so independent of dimension.

\section{Conclusion and outlook}\label{sec:conclusion}

In this work we have studied polydisperse two-dimensional (square lattice) and unknotted three-dimensional (simple cubic lattice) lattice embeddings of theta graphs using a combination of the Wang-Landau Monte Carlo method and local BFACF moves to randomly sample configurations. Our results point to non-trivial values of the entropic exponents, new novel exponents for the arm lengths and the separation of the vertices of degree three (branch points). We estimate the entropic exponent for the square lattice to be $\zeta_\text{sq} = -1.458 \pm 0.005$ whilst on the cubic lattice we find $\zeta_\text{cu} =-1.67\pm 0.01$. These are distinct from the values $\alpha-2$ ($-1.5$ and $-1.763$ respectively), which would result from taking self-avoiding polygons of length $n$ and inserting a small loop in one of $n$ possible positions.

For the "short" arm and sum of two shortest arms length exponent we find there is likely only one value and that is  $0.577 \pm 0.005$ on the square lattice and $0.775 \pm 0.01$ on the cubic lattice. For the mean squared distance between the branch points, we estimate exponents of $0.530 \pm 0.005$ on the square lattice and either $0.460 \pm 0.005$ or $0.475 \pm 0.005$, depending on the correction to scaling exponent assumed. 

We have also looked at the entropic exponents in the monodisperse cases and our results support the previous prediction of \cite{duplantier_polymer_1986,duplantier_statistical_1989}
in two dimensions. We note that our estimate for this exponent on cubic lattice is very close to the square lattice result and exact prediction: this calls for further examination. Considering future directions it would be interesting to know whether the exponents in the polydisperse case in two dimensions have exact values that can be predicted. Further work will need to be undertaken to understand the relationship of our result to those of knotted polygons.

While it is in principle possible to incorporate non-local moves like pivots into a Monte Carlo method for thetas, the vertices of degree 3 would make these quite challenging to implement. Even with only local BFACF-type moves, there are two fairly straightforward potential extensions of this work:
\begin{itemize}
    \item Other objects: tadpoles, dumbbells, 3-stars and other spatial graphs with vertices of degree 3 can easily be sampled using the same methodology.
    \item Other lattices: We expect that a similar set of moves to those of \cref{fig:sq_BFACF_theta} involving vertices of degree 3 can be implemented for the triangular and FCC lattices (in fact they are simpler).
\end{itemize}

Finally, to reiterate here in three dimensions we studied \emph{unknotted} theta-graphs. There have now been studies of how knots affect self-avoiding polygons and in this work how the theta topology modifies the scaling behaviour. It is not entirely clear whether different knot types of thetas would behave in the same manner as unknotted thetas, and so the interaction of knottedness with the theta topology would be of real interest. 

\section*{Acknowledgements}

We thank James Gleeson whose precursor work as part of their Master's thesis provided the background for this study. Financial support was provided by the Australian Research Council Discovery Project DP230100674. Computational support was provided by the University of Melbourne Research Computing Services. The authors are grateful for helpful conversations with Nathan Clisby, Stu Whittington, Chris Soteros, Andrew Rechnitzer and Tony Guttmann.

\printbibliography

@article{wang_landau_2001,
  title = {Efficient, Multiple-Range Random Walk Algorithm to Calculate the Density of States},
  author = {Wang, Fugao and Landau, D. P.},
  journal = {Phys. Rev. Lett.},
  volume = {86},
  number = {10},
  pages = {2050--2053},
  numpages = {0},
  year = {2001},
  publisher = {American Physical Society},
  doi = {10.1103/PhysRevLett.86.2050},
}

@article{jacobsen_growth_2016,
	title = {On the growth constant for square-lattice self-avoiding walks},
	volume = {49},
	doi = {10.1088/1751-8113/49/49/494004},
	pages = {494004},
	number = {49},
	journal = {J. Phys. A: Math. Theor.},
	author = {Jacobsen, Jesper Lykke and Scullard, Christian R. and Guttmann, Anthony J.},
	year = {2016},
}

@article{clisby_calculation_2013,
	title = {Calculation of the connective constant for self-avoiding walks via the pivot algorithm},
	volume = {46},
	doi = {10.1088/1751-8113/46/24/245001},
	pages = {245001},
	number = {24},
	journal = {J. Phys. A: Math. Theor.},
	author = {Clisby, Nathan},
	year = {2013},
}

@article{Arag_o_de_Carvalho_1983, 
    title={A new Monte-Carlo approach to the critical properties of self-avoiding random walks}, 
    volume={44}, 
    DOI={10.1051/jphys:01983004403032300}, 
    number={3}, 
    journal={J. Physique}, 
    publisher={EDP Sciences}, 
    author={Aragão de Carvalho, C. and Caracciolo, S.}, 
    year={1983}, 
    pages={323–331} 
}

@article{de_Carvalho_1983, 
    title={Polymers and $g|\varphi|^4$ theory in four dimensions}, 
    volume={215}, 
    DOI={10.1016/0550-3213(83)90213-4}, 
    number={2}, 
    journal={Nuclear Physics B}, 
    publisher={Elsevier BV}, 
    author={Aragão de Carvalho, C. and Caracciolo, S. and Fröhlich, J.},    
    year={1983}, 
    pages={209–248} 
}

@article{Berg_1981, 
    title={Random paths and random surfaces on a digital computer}, 
    volume={106}, 
    DOI={10.1016/0370-2693(81)90545-1}, 
    number={4}, 
    journal={Phys. Lett. B}, 
    publisher={Elsevier BV}, 
    author={Berg, B. and Foerster, D.}, 
    year={1981}, 
    pages={323–326} 
}

@article{janse_van_rensburg_bfacf_1991,
	title = {The {BFACF} algorithm and knotted polygons},
	volume = {24},
	doi = {10.1088/0305-4470/24/23/021},
	pages = {5553--5567},
	number = {23},
	journal = {J. Phys. A: Math. Gen.},
	author = {Janse van Rensburg, E. J. and Whittington, S. G.},
	year = {1991},
}

@article{janse_van_rensburg_bfacf-style_2011,
	title = {{BFACF}-style algorithms for polygons in the body-centered and face-centered cubic lattices},
	volume = {44},
	doi = {10.1088/1751-8113/44/16/165001},
	pages = {165001},
	number = {16},
	journal = {J. Phys. A: Math. Theor.},
	author = {Janse van Rensburg, E. J. and Rechnitzer, A.},
	year = {2011},
}

@mastersthesis{Tamaki2018,
   author = {Kanami Tamaki},
   shortauthor  = {Tamaki},
   school = {Saitama University},
   title = {Knots and Spatial Graphs in the Simple Cubic Lattice},
   year = {2018},
}

@article{bradly_force-induced_2019,
	title = {Force-induced desorption of 3-star polymers in two dimensions},
	volume = {52},
	doi = {10.1088/1751-8121/ab2af4},
	pages = {315002},
	number = {31},
	journal = {J. Phys. A: Math. Theor.},
	author = {Bradly, C J and Janse Van Rensburg, E J and Owczarek, A L and Whittington, S G},
	year = {2019},
}

@article{janse_van_rensburg_monte_2009,
	title = {Monte Carlo methods for the self-avoiding walk},
	volume = {42},
	doi = {10.1088/1751-8113/42/32/323001},
	pages = {323001},
	number = {32},
	journal = {J. Phys. A: Math. Theor.},
	author = {Janse van Rensburg, E. J.},
	year = {2009},
}

@article{clisby_new_2012,
	title = {A new transfer-matrix algorithm for exact enumerations: self-avoiding polygons on the square lattice},
	volume = {45},
	doi = {10.1088/1751-8113/45/11/115202},
	pages = {115202},
	number = {11},
	journal = {J. Phys. A: Math. Theor.},
	author = {Clisby, Nathan and Jensen, Iwan},
	year = {2012},
}

@article{moriuchi_enumeration_2009,
	title = {An enumeration of theta-curves with up to seven crossings},
	volume = {18},
	doi = {10.1142/S0218216509006884},
	pages = {167--197},
	number = {2},
	journal = {J. Knot Theory Ramifications},
	author = {Moriuchi, Hiromasa},
	year = {2009},
}

@collection{guttmann_polygons_2009,
	title = {Polygons, Polyominoes and Polycubes},
	series = {Lecture Notes in Physics},
	publisher = {Springer Netherlands},
	editor = {Guttmann, A. J.},
	year = {2009},
}

@article{caracciolo_correction--scaling_2005,
	title = {Correction-to-Scaling Exponents for Two-Dimensional Self-Avoiding Walks},
	volume = {120},
	doi = {10.1007/s10955-005-7004-3},
	pages = {1037--1100},
	number = {5},
	journal = {J. Stat. Phys.},
	author = {Caracciolo, Sergio and Guttmann, Anthony J. and Jensen, Iwan and Pelissetto, Andrea and Rogers, Andrew N. and Sokal, Alan D.},
	year = {2005},
}

@article{clisby_high-precision_2016,
	title = {High-precision estimate of the hydrodynamic radius for self-avoiding walks},
	volume = {94},
	doi = {10.1103/PhysRevE.94.052102},
	pages = {052102},
	number = {5},
	journal = {Phys. Rev. E},
	author = {Clisby, Nathan and Dünweg, Burkhard},
	year = {2016},
}

@article{kim_lattice_2022,
	title = {Lattice conformation of theta-curves accompanied with Brunnian property},
	volume = {55},
	doi = {10.1088/1751-8121/ac845a},
	pages = {435207},
	number = {43},
	journal = {J. Phys. A: Math. Theor.},
	publisher = {{IOP} Publishing},
	author = {Kim, Hyoungjun and Lee, Hwa Jeong and No, Sungjong and Oh, Seungsang and Yoo, Hyungkee},
	year = {2022},
}

@article{no_topological_2021,
	title = {Topological aspects of theta-curves in cubic lattice},
	volume = {54},
	doi = {10.1088/1751-8121/ac2ae9},
	pages = {455204},
	number = {45},
	journal = {J. Phys. A: Math. Theor.},
	publisher = {{IOP} Publishing},
	author = {No, Sungjong and Oh, Seungsang and Yoo, Hyungkee},
	year = {2021},
}

@article{sykes_counting_1961,
	title = {Some Counting Theorems in the Theory of the Ising Model and the Excluded Volume Problem},
	volume = {2},
	doi = {10.1063/1.1724212},
	pages = {52--62},
	number = {1},
	journal = {J. Math. Phys.},
	author = {Sykes, M. F.},
	year = {1961},
}

@article{guttmann_two-dimensional_1978,
	title = {Two-dimensional lattice embeddings of connected graphs of cyclomatic index two},
	volume = {11},
	doi = {10.1088/0305-4470/11/4/013},
	pages = {721--729},
	number = {4},
	journal = {J. Phys. A: Math. Gen.},
	publisher = {{IOP} Publishing},
	author = {Guttmann, A. J. and Whittington, S. G.},
	year = {1978},
}

@article{guttmann_self-avoiding_2004,
	title = {Self-avoiding walks and trails on the $3.12^2$ lattice},
	volume = {38},
	doi = {10.1088/0305-4470/38/3/002},
	pages = {543},
	number = {3},
	journal = {J. Phys. A: Math. Gen.},
	author = {Guttmann, Anthony J and Parviainen, Robert and Rechnitzer, Andrew},
	year = {2004},
}

@misc{guttmann_private_comm,
    author = {Guttmann, Anthony J.},
    note = {Private communication},
    year = {2021}
}

@article{duplantier_polymer_1986,
	title = {Polymer Network of Fixed Topology: Renormalization, Exact Critical Exponent $\gamma$ in Two Dimensions, and $d=4-\varepsilon$},
	volume = {57},
	doi = {10.1103/PhysRevLett.57.2332},
	pages = {2332--2332},
	number = {18},
	journal = {Phys. Rev. Lett.},
	author = {Duplantier, Bertrand},
	year = {1986},
}

@article{janse_van_rensburg_knot_1990,
	title = {The knot probability in lattice polygons},
	volume = {23},
	doi = {10.1088/0305-4470/23/15/028},
	pages = {3573},
	number = {15},
	journal = {J. Phys. A: Math. Gen.},
	author = {Janse van Rensburg, E. J. and Whittington, S. G.},
	year = {1990},
}

@article{janse_van_rensburg_thoughts_2008,
	title = {Thoughts on lattice knot statistics},
	volume = {45},
	doi = {10.1007/s10910-008-9364-9},
	pages = {7},
	number = {1},
	journal = {J. Math. Chem.},
	author = {Janse van Rensburg, E. J.},
	year = {2008},
}

@article{clisby_accurate_2010,
	title = {Accurate Estimate of the Critical Exponent $\nu$ for Self-Avoiding Walks via a Fast Implementation of the Pivot Algorithm},
	volume = {104},
	doi = {10.1103/PhysRevLett.104.055702},
	pages = {055702},
	number = {5},
	journal = {Phys. Rev. Lett.},
	author = {Clisby, Nathan},
	year = {2010},
}

@article{duplantier_statistical_1989,
	title = {Statistical mechanics of polymer networks of any topology},
	volume = {54},
	doi = {10.1007/BF01019770},
	pages = {581--680},
	number = {3},
	journal = {J. Stat. Phys.},
	author = {Duplantier, Bertrand},
	year = {1989},
}

@article{frisch_chemical_1961,
	title = {Chemical Topology},
	volume = {83},
	url = {https://doi.org/10.1021%2Fja01479a015},
	doi = {10.1021/ja01479a015},
	pages = {3789--3795},
	number = {18},
	journal = {J. Amer. Chem. Soc.},
	publisher = {American Chemical Society ({ACS})},
	author = {Frisch, H. L. and Wasserman, E.},
	year = {1961},
}

@inproceedings{delbruck_mathematical_1962,
	location = {Providence, {RI}},
	title = {Mathematical Problems in the Biological Sciences},
	volume = {14},
	doi = {https://doi.org/10.1090/psapm/014},
	pages = {55--63},
	booktitle = {Proceedings of the Symposium on Applied Mathematics},
	publisher = {American Mathematical Society},
	author = {Delbruck, Max and Fuller, F. Brock},
	date = {1962},
}

@article{tubiana_topology_2024,
	title = {Topology in soft and biological matter},
	volume = {1075},
	doi = {10.1016/j.physrep.2024.04.002},
	pages = {1--137},
	journal = {Physics Reports},
	author = {Tubiana, Luca and Alexander, Gareth P. and Barbensi, Agnese and Buck, Dorothy and Cartwright, Julyan H. E. and Chwastyk, Mateusz and Cieplak, Marek and Coluzza, Ivan and Čopar, Simon and Craik, David J. and Di Stefano, Marco and Everaers, Ralf and Faísca, Patrícia F. N. and Ferrari, Franco and Giacometti, Achille and Goundaroulis, Dimos and Haglund, Ellinor and Hou, Ya-Ming and Ilieva, Nevena and Jackson, Sophie E. and Japaridze, Aleksandre and Kaplan, Noam and Klotz, Alexander R. and Li, Hongbin and Likos, Christos N. and Locatelli, Emanuele and López-León, Teresa and Machon, Thomas and Micheletti, Cristian and Michieletto, Davide and Niemi, Antti and Niemyska, Wanda and Niewieczerzal, Szymon and Nitti, Francesco and Orlandini, Enzo and Pasquali, Samuela and Perlinska, Agata P. and Podgornik, Rudolf and Potestio, Raffaello and Pugno, Nicola M. and Ravnik, Miha and Ricca, Renzo and Rohwer, Christian M. and Rosa, Angelo and Smrek, Jan and Souslov, Anton and Stasiak, Andrzej and Steer, Danièle and Sułkowska, Joanna and Sułkowski, Piotr and Sumners, De Witt L. and Svaneborg, Carsten and Szymczak, Piotr and Tarenzi, Thomas and Travasso, Rui and Virnau, Peter and Vlassopoulos, Dimitris and Ziherl, Primož and Žumer, Slobodan},
	year = {2024},
}

@article{marcone_size_2007,
	title = {Size of knots in ring polymers},
	volume = {75},
	doi = {10.1103/PhysRevE.75.041105},
	pages = {041105},
	number = {4},
	journal = {Phys. Rev. E},
	author = {Marcone, B. and Orlandini, E. and Stella, A. L. and Zonta, F.},
	year = {2007},
}

@article{orlandini_size_2009,
	title = {The size of knots in polymers},
	volume = {6},
	doi = {10.1088/1478-3975/6/2/025012},
	pages = {025012},
	number = {2},
	journal = {Phys. Biol.},
	publisher = {{IOP} Publishing},
	author = {Orlandini, Enzo and Stella, Attilio L. and Vanderzande, Carlo},
	year = {2009},
}

@article{tezuka_synthesis_2003,
	title = {Synthesis of $\theta$-Shaped Poly({THF}) by Electrostatic Self-Assembly and Covalent Fixation with Three-Armed Star Telechelics Having Cyclic Ammonium Salt Groups},
	volume = {36},
	doi = {10.1021/ma0209850},
	pages = {65--70},
	number = {1},
	journal = {Macromolecules},
	publisher = {American Chemical Society},
	author = {Tezuka, Yasuyuki and Tsuchitani, Akiko and Yoshioka, Yuka and Oike, Hideaki},
	year = {2003},
}

@article{gu_polymer_networks_2020,
author = {Gu, Yuwei and Zhao, Julia and Johnson, Jeremiah A.},
title = {Polymer Networks: From Plastics and Gels to Porous Frameworks},
journal = {Angewandte Chemie International Edition},
volume = {59},
number = {13},
pages = {5022-5049},
doi = {10.1002/anie.201902900},
year = {2020}
}

@article{uehara_statistical_2018,
	title = {Statistical properties of multi-theta polymer chains},
	volume = {51},
	doi = {10.1088/1751-8121/aaae2d},
	pages = {134001},
	number = {13},
	journal = {J. Phys. A: Math. Theor.},
	publisher = {{IOP} Publishing},
	author = {Uehara, Erica and Deguchi, Tetsuo},
	year = {2018},
}

@article{janse_van_rensburg_exponential_2024,
	title = {Exponential growth rate of lattice comb polymers},
	volume = {57},
	doi = {10.1088/1751-8121/ad8a2d},
	pages = {485002},
	number = {48},
	journal = {J. Phys. A: Math. Theor.},
	author = {Janse van Rensburg, E J and Whittington, S G},
	year = {2024},
}

@article{guttmann_limiting_1973,
	title = {Limiting ring closure probability index for the self avoiding random walk problem},
	volume = {6},
	doi = {10.1088/0022-3719/6/6/009},
	pages = {945},
	number = {6},
	journal = {J. Phys. C: Solid State Phys.},
	author = {Guttmann, A. J. and Sykes, M. F.},
	year = {1973},
}

@article{rensburg_dimensions_1991,
	title = {The dimensions of knotted polygons},
	volume = {24},
	doi = {10.1088/0305-4470/24/16/028},
	pages = {3935},
	number = {16},
	journal = {J. Phys. A: Math. Gen.},
	author = {Janse van Rensburg, E. J.  and Whittington, S. G.},
	year = {1991},
}

@article{janse_van_rensburg_generalized_2012,
	title = {Generalized atmospheric sampling of knotted polygons},
	doi = {10.1142/S0218216511009170},
	journal = {J. Knot Theory Ramifications},
	author = {Janse van Rensburg, E. J. and Rechnitzer, A.},
	year = {2012},
}

@article{landau_new_2004,
	title = {A new approach to Monte Carlo simulations in statistical physics: Wang-Landau sampling},
	volume = {72},
	doi = {10.1119/1.1707017},
	pages = {1294--1302},
	number = {10},
	journal = {Am. J. Phys.},
	author = {Landau, D. P. and Tsai, Shan-Ho and Exler, M.},
	year = {2004},
}

@article{belardinelli_fast_2007,
	title = {Fast algorithm to calculate density of states},
	volume = {75},
	doi = {10.1103/PhysRevE.75.046701},
	pages = {046701},
	number = {4},
	journal = {Phys. Rev. E},
	author = {Belardinelli, R. E. and Pereyra, V. D.},
	year = {2007},
}

@article{farago_pulling_2002,
	title = {Pulling knotted polymers},
	volume = {60},
	doi = {10.1209/epl/i2002-00317-0},
	pages = {53},
	number = {1},
	journal = {Europhys Lett.},
	author = {Farago, O. and Kantor, Y. and Kardar, M.},
	year = {2002},
}

@article{virnau_knots_2005,
	title = {Knots in Globule and Coil Phases of a Model Polyethylene},
	volume = {127},
	doi = {10.1021/ja052438a},
	pages = {15102--15106},
	number = {43},
	shortjournal = {J. Am. Chem. Soc.},
	author = {Virnau, Peter and Kantor, Yacov and Kardar, Mehran},
	date = {2005},
}

@article{mansfield_properties_2010,
	title = {Properties of knotted ring polymers. I. Equilibrium dimensions},
	volume = {133},
	doi = {10.1063/1.3457160},
	pages = {044903},
	number = {4},
	journal = {J. Chem. Phys.},
	author = {Mansfield, Marc L. and Douglas, Jack F.},
	year = {2010},
}

@article{tubiana_spontaneous_2013,
	title = {Spontaneous Knotting and Unknotting of Flexible Linear Polymers: Equilibrium and Kinetic Aspects},
	volume = {46},
	doi = {10.1021/ma4002963},
	shorttitle = {Spontaneous Knotting and Unknotting of Flexible Linear Polymers},
	pages = {3669--3678},
	number = {9},
	journal = {Macromolecules},
	author = {Tubiana, L. and Rosa, A. and Fragiacomo, F. and Micheletti, C.},
	date = {2013},
}

@article{Atisattapong_2021, 
    title={Wang–Landau sampling for estimation of the reliability of physical networks}, 
    volume={262}, 
    DOI={10.1016/j.cpc.2021.107831}, 
    journal={Computer Physics Communications}, 
    author={Atisattapong, Wanyok and Marupanthorn, Pasin}, 
    year={2021}, 
    pages={107831} 
}

@article{LI2007524,
    title = {Numerical integration using Wang–Landau sampling},
    journal = {Computer Physics Communications},
    volume = {177},
    number = {6},
    pages = {524-529},
    year = {2007},
    doi = {10.1016/j.cpc.2007.06.001},
    author = {Y.W. Li and T. Wüst and D.P. Landau and H.Q. Lin},
}

@article{taylor_phase_2009,
	title = {Phase transitions of a single polymer chain: A Wang–Landau simulation study},
	volume = {131},
	doi = {10.1063/1.3227751},
	shorttitle = {Phase transitions of a single polymer chain},
	pages = {114907},
	number = {11},
	journal = {J. Chem. Phys.},
	author = {Taylor, Mark P. and Paul, Wolfgang and Binder, Kurt},
	year = {2009},
}

@article{wust_versatile_2009,
	title = {Versatile Approach to Access the Low Temperature Thermodynamics of Lattice Polymers and Proteins},
	volume = {102},
	doi = {10.1103/PhysRevLett.102.178101},
	pages = {178101},
	number = {17},
	journal = {Phys. Rev. Lett.},
	author = {Wüst, Thomas and Landau, David P.},
	date = {2009-04-29},
}

@article{ZHAN2008339,
    title = {A parallel implementation of the Wang–Landau algorithm},
    journal = {Computer Physics Communications},
    volume = {179},
    number = {5},
    pages = {339-344},
    year = {2008},
    issn = {0010-4655},
    doi = {10.1016/j.cpc.2008.04.002},
    author = {Lixin Zhan},
}

@article{YIN20121568,
    title = {Massively parallel Wang–Landau sampling on multiple GPUs},
    journal = {Computer Physics Communications},
    volume = {183},
    number = {8},
    pages = {1568-1573},
    year = {2012},
    doi = {10.1016/j.cpc.2012.02.023},
    author = {Junqi Yin and D.P. Landau},
}

@article{malakis_wanglandau_2004,
	title = {On the Wang–Landau method using the N-fold way},
	volume = {15},
	doi = {10.1142/S0129183104006182},
	pages = {729--740},
	number = {5},
	journal = {Int. J. Mod. Phys. C},
	author = {Malakis, A. and Martinos, S. S. and Hadjiagapiou, I. A. and Peratzakis, A. S.},
	year = {2004},
}

@book{madras_self-avoiding_1996,
	location = {Boston},
	title = {The Self-Avoiding Walk},
	isbn = {978-0-8176-3891-7 978-3-7643-3891-6 978-0-8176-3589-3},
	publisher = {Birkhäuser},
	author = {Madras, Neal and Slade, Gordon},
	year = {1996},
}

@Book{DeGennes1979,
  author    = {{De Gennes}, P.G.},
  title     = {Scaling Concepts in Polymer Physics},
  isbn      = {9780801412035},
  publisher = {Cornell University Press},
  lccn      = {lc78021314},
  year      = {1979},
}

@book{janse_van_rensburg_statistical_2015,
	location = {Oxford, New York},
	edition = {Second Edition},
	title = {The Statistical Mechanics of Interacting Walks, Polygons, Animals and Vesicles},
	series = {Oxford Lecture Series in Mathematics and Its Applications},
	pagetotal = {640},
	publisher = {Oxford University Press},
	author = {Janse van Rensburg, E. J.},
	year = {2015},
}

@article{baiesi_entropic_2010,
	title = {The entropic cost to tie a knot},
	volume = {2010},
	doi = {10.1088/1742-5468/2010/06/P06012},
	pages = {P06012},
	number = {6},
	journal = {J. Stat. Mech.},
	author = {Baiesi, M and Orlandini, E and Stella, A L},
	year = {2010},
}

@article{sumners_knots_1988,
	title = {Knots in self-avoiding walks},
	volume = {21},
	doi = {10.1088/0305-4470/21/7/030},
	pages = {1689},
	number = {7},
	journal = {J. Phys. A: Math. Gen.},
	author = {Sumners, D. W. and Whittington, S. G.},
	year = {1988},
}

@article{janse_van_rensburg_universality_2011,
	title = {On the universality of knot probability ratios},
	volume = {44},
	doi = {10.1088/1751-8113/44/16/162002},
	pages = {162002},
	number = {16},
	journal = {J. Phys. A: Math. Theor.},
	author = {Janse van Rensburg, E. J. and Rechnitzer, A.},
	year = {2011},
}

@article{beaton_entanglement_2026,
	title = {Entanglement statistics of polymers in a lattice tube and unknotting of 4-plats},
	volume = {379},
	doi = {10.1016/j.dam.2025.08.042},
	pages = {242--271},
	journal = {Discrete Applied Mathematics},
	author = {Beaton, Nicholas R. and Ishihara, Kai and Atapour, Mahshid and Eng, Jeremy W. and Vazquez, Mariel and Shimokawa, Koya and Soteros, Christine E.},
	year = {2026},
}

@mastersthesis{gleeson_thesis_2024,
    author = {James Gleeson},
    title = {Monte Carlo Enumeration of Topological Polymers},
    school = {University of Melbourne},
    year = {2024},
}

@article{dabrowski-tumanski_theta-curves_2024,
	title = {Theta-curves in proteins},
	volume = {33},
	doi = {10.1002/pro.5133},
	pages = {e5133},
	number = {9},
	journaltitle = {Protein Science},
	author = {Dabrowski-Tumanski, Pawel and Goundaroulis, Dimos and Stasiak, Andrzej and Rawdon, Eric J. and Sulkowska, Joanna I.},
	year = {2024},
}

@misc{whittington_private_comm,
    author = {Whittington, Stuart G.},
    note = {Private communication},
    year = {2026}
}

\end{document}